\documentclass[preprint2]{aastex}

\def\gsim{\mathrel{\hbox{\rlap{\hbox{\lower4pt\hbox{$\sim$}}}\hbox{$>$}}}}
\def\lsim{\mathrel{\hbox{\rlap{\hbox{\lower4pt\hbox{$\sim$}}}\hbox{$<$}}}}

\def\ax{{\it AEGIS-X}}
\def\chandra{{\it Chandra}}

\def\fxfopt{{$F_{\rm X}/F_{\rm opt}$}}

\slugcomment{Resubmitted to ApJ Supp}

\shorttitle{AEGIS-X Deep Catalogue}
\shortauthors{K. Nandra et~al.}

\begin{document}


\title{{\it AEGIS-X:} Deep Chandra imaging of the Central Groth Strip}


\author{K. Nandra\altaffilmark{1,2},  
E.S. Laird\altaffilmark{2}, 
J.A. Aird\altaffilmark{3,4},
M. Salvato\altaffilmark{1},
A. Georgakakis\altaffilmark{1},
G. Barro\altaffilmark{5,6},
P.G. Perez-Gonzalez\altaffilmark{6},
P. Barmby\altaffilmark{7}, 
R.-R. Chary\altaffilmark{8},
A. Coil\altaffilmark{9},
M.C. Cooper\altaffilmark{10},
M. Davis\altaffilmark{11}, 
M. Dickinson\altaffilmark{12},
S.M. Faber\altaffilmark{5}, 
G.G. Fazio\altaffilmark{13}, 
P. Guhathakurta\altaffilmark{5},
S. Gwyn\altaffilmark{14},
L.-T. Hsu\altaffilmark{1},
J.-S. Huang\altaffilmark{10},
R.J. Ivison\altaffilmark{15},
D.C. Koo\altaffilmark{5}, 
J.A. Newman\altaffilmark{16},
C. Rangel\altaffilmark{2},
T. Yamada\altaffilmark{17},
C. Willmer\altaffilmark{18}
}
\altaffiltext{1}{Max Planck Institute for Extraterrestrial Physics, Giessenbachstrasse, 85741 Garching, Germany}
\altaffiltext{2}{Astrophysics Group, Blackett Laboratory, Imperial College London, London SW7 2AZ, UK}
\altaffiltext{3}{Center for Astrophysics and Space Science, University of California, San Diego, CA 92093, USA}
\altaffiltext{4}{University of Durham, UK}
\altaffiltext{5}{UCO/Lick Observatory, Department of Astronomy and Astrophysics, University of California, Santa Cruz, CA 95064}
\altaffiltext{6}{{Departamento de Astrof\'{\i}sica, Facultad de CC. F\'{\i}sicas, Universidad Complutense de Madrid, E-28040 Madrid, Spain}}
\altaffiltext{7}{Department of Physics and Astronomy, University of Western Ontario, London, Ontario N6A 3K7, Canada}
\altaffiltext{8}{MS220-6, Caltech, Pasadena, CA 91125, USA}
\altaffiltext{9}{Department of Physics, Center for Astrophysics and Space Sciences, University of California at San Diego, 9500 Gilman Dr., La Jolla, San Diego, CA 92093}
\altaffiltext{10}{Center for Galaxy Evolution, Department of Physics and Astronomy, University of California, Irvine, CA 92697, USA}
\altaffiltext{11}{Department of Astronomy, University of California, Berkeley, CA 94720}
\altaffiltext{12}{NOAO, Tucson, AZ 85719, USA}
\altaffiltext{13}{Harvard-Smithsonian Center for Astrophysics, Cambridge, MA 02138}
\altaffiltext{14}{CADC, HIA, Victoria, Canada}
\altaffiltext{15}{Institute for Astronomy, University of Edinburgh, Royal Observatory, Blackford Hill, Edinburgh EH9 3HJ, UK}
\altaffiltext{16}{University of Pittsburgh and Pitt-PACC, 3941 O'Hara St., Pittsburgh, PA 15260}
\altaffiltext{17}{Tohoku University, Aramaki, Aoba, Sendai 9808578, Japan}
\altaffiltext{18}{Steward Observatory, University of Arizona, 933 North Cherry Avenue, Tucson, AZ 85721, USA}


\begin{abstract}

We present the results of deep \chandra\ imaging of the central region of the Extended Groth Strip, the AEGIS-X Deep (AEGIS-XD) survey. When combined with previous \chandra\ observations of a wider area of the strip, AEGIS-X Wide (AEGIS-XW; Laird et~al. 2009), these provide data to a nominal exposure depth of 800ks in the three central ACIS-I fields, a region of approximately $0.29$~deg$^{2}$. This is currently the third deepest X-ray survey in existence, a factor $\sim 2-3$ shallower than the Chandra Deep Fields (CDFs) but over an area $\sim 3$ times greater than each CDF. We present a catalogue of 937 point sources detected in the deep \chandra\ observations. We present identifications of our X-ray sources from deep ground-based, Spitzer, GALEX and HST imaging. Using a likelihood ratio analysis, we associate multi band counterparts for 929/937 of our X-ray sources, with an estimated 95~\% reliability, making the identification completeness approximately 94~\% in a statistical sense.  Reliable spectroscopic redshifts for 353 of our X-ray sources are provided predominantly from Keck (DEEP2/3) and MMT Hectospec, so the current spectroscopic completeness is $\sim 38$~per cent. For the remainder of the X-ray sources, we compute photometric redshifts based on multi-band photometry in up to 35  bands from the UV to mid-IR. Particular attention is given to the fact that the vast majority the X-ray sources are AGN and require hybrid templates. Our photometric redshifts have mean accuracy of  $\sigma=0.04$ and an outlier fraction of approximately 5\%, reaching $\sigma=0.03$ with less than 4\% outliers in the area covered by CANDELS . The X-ray, multi-wavelength photometry and redshift catalogues are made publicly available. 
\end{abstract}




\section{Introduction}
\label{sec:intro}

Deep X-ray surveys trace the accretion history of the universe, offering a highly efficient method of pinpointing growing black holes in galaxies over a wide range of redshifts. {\it Chandra} and {\it XMM-Newton} surveys have yielded samples of Active Galactic Nuclei (AGN) capable of characterising the evolution of accretion power in the Universe, via the X-ray luminosity function (XLF; e.g. Hasinger et~al. 2005; Barger et~al. 2005; Silverman et~al. 2008; Aird et~al. 2010; Ueda et~al. 2014). These surveys have also been highly influential in broadening our understanding of the co-evolution of supermassive black holes and galaxies via the characterisation of various host properties of AGN. The extraordinary sensitivity of the current generation of X-ray observatories, particularly {\it Chandra}, to point-like X-ray sources has transformed such investigations by revealing large populations of AGN in galaxies where the accretion activity in other wavebands is either obscured, or overwhelmed by host galaxy light (e.g. Brandt \& Hasinger 2005). 

The deepest X-ray surveys thus far are the Chandra Deep Fields (CDFs; Giacconi et~al. 2002; Alexander et~al. 2003; Luo et~al. 2010; Xue et~al. 2011). While these reach extremely faint flux levels they can only do so over relatively small areas. Complementary large-area, but shallower surveys have therefore been performed, such as XBootes (Murray et~al. 2005), XMM-LSS (Pierre et~al. 2007), XMM-COSMOS (Hasinger et~al. 2007) and {\it Chandra} COSMOS (Elvis et~al. 2009).  

Determination of the accretion history and its relationship to galaxy evolution cannot be achieved using X-ray data alone. For example, a basic requirement is also to determine the redshifts of the X-ray sources to determine their luminosities and evolution. This has proved surprisingly difficult, for a number of reasons. Firstly, the depth of current X-ray observations is such that the multiwavelength counterparts of the X-ray sources are often extremely faint. This makes even the identification e.g. of optical or NIR counterparts challenging. Determining their redshifts is even more difficult, because the vast majority are too faint for spectroscopic identification. Despite major efforts, the spectroscopic completeness of the deepest X-ray samples are $<50$~\% (e.g. Szokoly et~al. 2004; Trouille et~al. 2008). Photometric redshifts can be used to mitigate this spectroscopic incompleteness, but require very deep data in as many bands as possible. This can be difficult to acquire in wide fields. Also, because the vast majority of X-ray point sources in deep surveys are AGN, special consideration is also required to yield accurate photo-z's (Salvato et~al. 2009, 2011). 

One area of the sky which benefits from deep multi wavelength coverage is  the Extended Groth Strip (EGS), a region of 0.25 x 2' centered at approximately $\alpha$ = 14$^h$ \ 18$^m$, $\delta=52^{\circ} \ 00^{\prime}$. Deep observations of the EGS have been performed using ground and space-based observatories, many of them as part of the AEGIS multiwavelength project (Davis et~al. 2007). This includes X-ray imaging with {\it Chandra}/ACIS, which covers the entire EGS to a nominal depth of 200ks, which we henceforth designate the AEGIS-X Wide (AEGIS-XW) survey  (Nandra et~al. 2005; Laird et~al. 2009 hereafter L09). These observations have been effective in helping characterizing the accretion history to relatively high redshifts (Aird et~al. 2008; 2010). In combination with the AEGIS multi wavelength data they have also provided new insights into the relationship of AGN with their host galaxies (e.g. Nandra et~al. 2007; Pierce et~al. 2007; Bundy et~al. 2008; Georgakakis et~al. 2008c, 2009) and large scale structure environments  (e.g. Georgakakis et~al. 2007, 2008a Coil et~al. 2009). 

In the current paper we present deeper {\it Chandra} imaging (800ks nominal depth) of the central EGS region, hereafter the AEGIS-X Deep (AEGIS-XD) survey. The AEGIS-XW data are sufficient to detect essentially the full population of X-ray emitting AGN to $z\sim3$ (Aird et~al. 2008). Understanding the evolution above this redshift is important, as above $z=3$ the luminous QSO population is known to decline rapidly (e.g Wall et~al. 2005). Whether this also applies to the obscured and more moderate luminosity AGN populations probed in X-ray surveys is an open question Brusa et~al., 2009; Civano et~al., 2011; Vito et~al. 2013). The CDFs reach sufficient depths to detect such sources at $z>3$ (e.g. Fiore et~al. 2012), but they do not cover sufficient area to yield samples of sufficient size to fully characterize the total accretion power at the redshifts of interest. A further advantage of deeper X-ray data is improved characterization of the X-ray spectral properties of deep field sources, and particularly the obscuration properties which may be strongly linked to SMBH-galaxy co-evolution scenarios (e.g. Hopkins et~al. 2005). Because most {\it Chandra} surveys are severely photon-starved, this is very challenging without long exposures. 

The AEGIS-XD data occupy a unique region of parameter space, being larger in area than each CDF by a factor ~3, but of sufficient depth to probe the high redshift and obscured X-ray source populations of interest. The AEGIS-XW data were designed to probe down to Seyfert level AGN activity ($L_{\rm X} \sim 10^{43}$~erg s$^{-1}$ at $z\sim3$ in the soft X-ray (0.5-2 keV) band. The new AEGIS-XD observations are sensitive to these luminosities at $z\sim5$, in principle with sufficient area to extend our understanding of the evolution of the X-ray luminosity function to at least that redshift. At the same time, in the harder X-ray band (2-10 keV), the AEGIS-XD images are sensitive to these kind of Seyfert level luminosities at $z\sim2$, even for sources absorbed by a column density of $N_{\rm H} = 10^{24}$~cm$^{-2}$. At these flux limits, sources with even higher column densities may be detected via their scattered light in either or both bands, via consideration of their X-ray spectra (e.g. Tozzi et~al. 2006; Brightman \& Ueda 2012). When combined with the exceptional multiwavelength data in AEGIS, the AEGIS-XD data therefore provide a unique resource to trace the growth of supermassive black holes, and its influence on galaxies, over a major fraction of cosmic time. 

In the present paper we describe the basic observational parameters and analysis of the AEGIS-XD  data, deriving an X-ray point source catalogue, a multi wavelength photometric catalog of the X-ray source counterparts, and a redshift catalogue. A companion paper presenting an extended source catalogue based on the same deep {\it Chandra} data has been presented in Erfanianfar et~al. (2013). An X-ray point source catalogue based on almost the same {\it Chandra} data used here has also been previously presented by Goulding et~al. (2012; hereafter G12). The present paper employs a different data reduction and source detection procedure, as well as a different technique for association of optical and near-infrared counterparts to the X-ray sources. Comparisons with this previous work are given in Sections~\ref{g12_xray} and \ref{g12_opt}. In addition, we provide a catalogue of spectroscopic and photometric redshifts for the X-ray source counterparts. The paper is structured as follows: Section 2 describes the X-ray observations and data reduction; Section 3 describes the source detection and photometry, which comprises an updated version of the AEGIS-XW methodology presented earlier by L09. The point source catalogue and sensitivity maps are presented in Section 4. Section 5 describes the multi-wavelength identifications and multi band photometry, the latter based on the methodology of the {\it Rainbow} database (Perez-Gonzalez et~al. 2008, Barro et~al. 2011a, 2011b). In Section 6 we present redshift estimates of the sources, including accurate photometric redshifts accounting for the AGN nature of the sources using the methods of Salvato et~al. (2011). A summary of our results is given in Section 7. 

\begin{figure*}
\begin{center}
{\scalebox{1.0}
{\includegraphics[trim=100 200 100 200,clip]{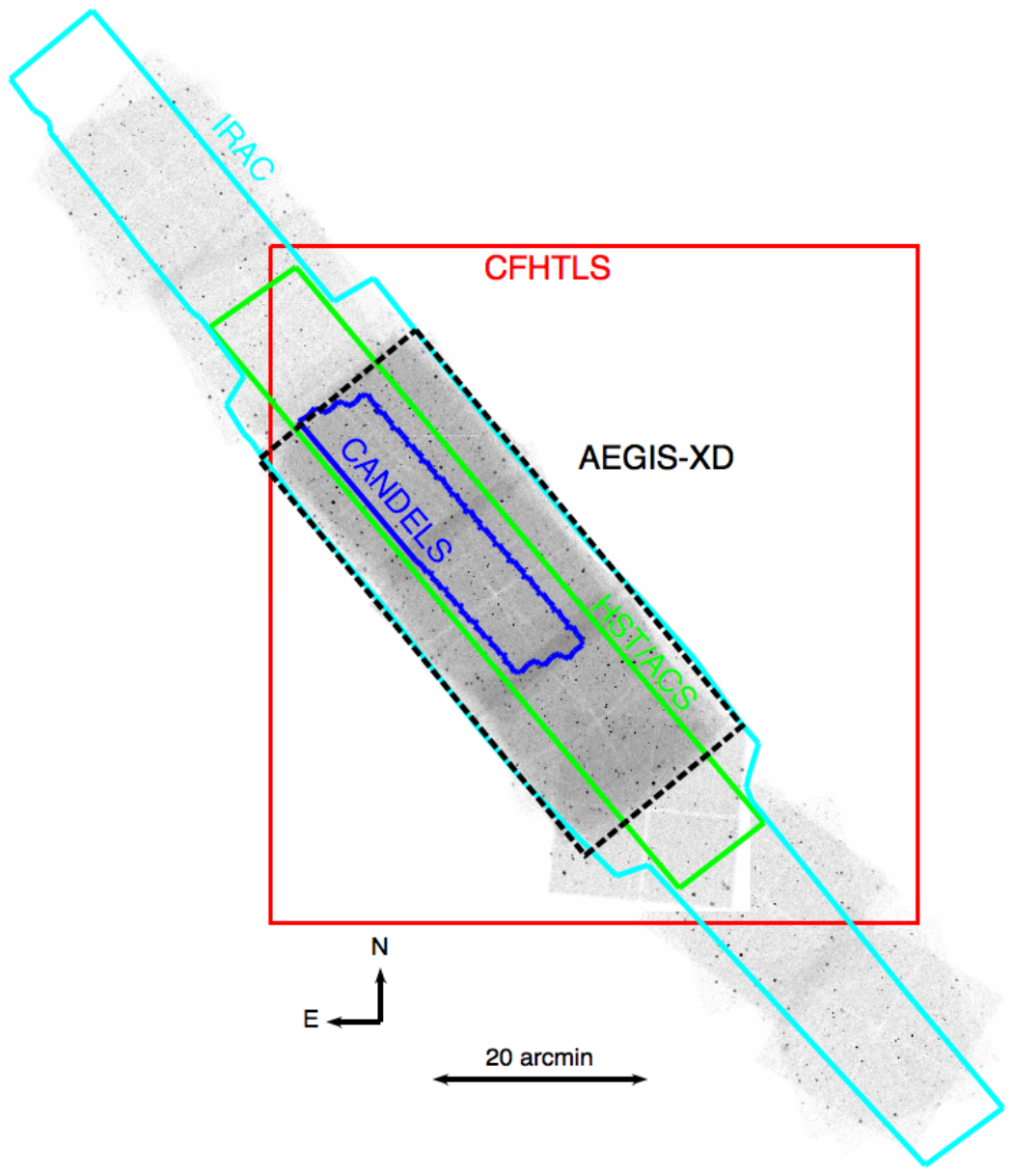}
}}
\caption{Layout of the AEGIS field showing the location of the {\it Chandra} X-ray imaging, and a subset of the multi wavelength coverage. The 200ks AEGIS-XW area (L09) is show as the greyscale image. The deeper 800ks coverage is contained in the area delineated within the thick black lines.
\label{fig:layout}}
\end{center}
\end{figure*}



\begin{deluxetable}{cccccccc}
\tabletypesize{\tiny}
\tablecaption{Observation Log \label{tab:obs}}
\tablewidth{0pt}
\tablehead{
\colhead{Field\tablenotemark{a}} & 
\colhead{ObsID\tablenotemark{b}} & 
 \colhead{RA\tablenotemark{c}} & 
\colhead{DEC\tablenotemark{c}} & 
\colhead{Start Time\tablenotemark{d}} & 
\colhead{On time\tablenotemark{e}} &
\colhead{Exposure\tablenotemark{f}} &
\colhead{Roll Angle\tablenotemark{g}} \\
\colhead{Name} & 
\colhead{} &
\colhead{(J2000)} & 
\colhead{(J2000)} & 
\colhead{(UT)} &  
\colhead{(ks)} & 
\colhead{(ks)} &
\colhead{(deg)} 
}
\startdata

AEGIS-1 &      9450&      14      20       17.24&+53      00       34.22&2007-12-11 04:24:07&       28.91&      28.78 &       40.2 \\ 
AEGIS-1 &      9451&      14      20       17.23&+53      00      34.23&2007-12-16 10:52:06&       25.38&       25.21 &       40.2 \\ 
AEGIS-1 &      9452&      14      20       16.80&+53      00      36.20&2007-12-18 05:45:49&       13.56&       13.29 &       48.7 \\ 
AEGIS-1 &      9453&      14      20       14.22&+52      59       43.05&2008-06-15 21:28:03&       44.75&      44.64 &      229.2 \\ 
AEGIS-1 &      9720&      14      20       14.21&+52      59       42.96&2008-06-17 05:14:02&       28.14&      27.74 &      229.2 \\ 
AEGIS-1 &      9721&      14      20       13.78&+52      59       45.13&2008-06-12 08:09:14&       16.55&      16.55 &      220.2 \\ 
AEGIS-1 &      9722&      14      20       13.76&+52      59       45.19&2008-06-13 07:02:28&       20.01&      19.89 &      220.2 \\ 
AEGIS-1 &      9723&      14      20       14.22&+52      59       43.00&2008-06-18 13:42:40&       34.54&      34.47 &      229.2 \\ 
AEGIS-1 &      9724&      14      20       16.82&+53      00      36.22&2007-12-22 13:37:26&       14.09&       14.08 &       48.7 \\ 
AEGIS-1 &      9725&      14      20       12.35&+53      00      16.04&2008-03-31 05:21:42&       31.13&       31.13 &      150.2 \\ 
AEGIS-1 &      9726&      14      20       13.77&+52      59       45.15&2008-06-05 08:45:04&       39.63&      39.62 &      220.2 \\ 
AEGIS-1 &      9793&      14      20       16.81&+53      00      36.27&2007-12-19 02:53:51&       24.08&       23.83 &       48.7 \\ 
AEGIS-1 &      9794&      14      20       16.82&+53      00      36.13&2007-12-20 04:27:59&       10.34&       10.03 &       48.7 \\ 
AEGIS-1 &      9795&      14      20       16.82&+53      00      36.32&2007-12-20 21:36:20&        8.91&        8.91 &       48.7 \\ 
AEGIS-1 &      9796&      14      20       16.81&+53      00      36.27&2007-12-21 20:28:33&       16.33&       16.33 &       48.7 \\ 
AEGIS-1 &      9797&      14      20       16.82&+53      00      36.38&2007-12-23 13:12:28&       12.63&       12.15 &       48.7 \\ 
AEGIS-1 &      9842&      14      20       12.35&+53      00      16.01&2008-04-02 21:01:59&       30.45&       30.44 &      150.2 \\ 
AEGIS-1 &      9843&      14      20       12.34&+53      00      16.13&2008-04-02 01:11:09&       15.32&       13.48 &      150.2 \\ 
AEGIS-1 &      9844&      14      20       12.35&+53      00      15.94&2008-04-05 13:07:54&       19.78&       19.78 &      150.2 \\ 
AEGIS-1 &      9863&      14      20       13.76&+52      59       45.14&2008-06-07 00:33:47&       22.01&      22.01 &      220.2 \\ 
AEGIS-1 &      9866&      14      20       13.77&+52      59       45.16&2008-06-03 22:43:14&       25.83&      25.83 &      220.2 \\ 
AEGIS-1 &      9870&      14      20       13.77&+52      59       44.99&2008-06-10 15:11:23&       11.08&      11.00 &      220.2 \\ 
AEGIS-1 &      9873&      14      20       13.77&+52      59       45.16&2008-06-11 14:22:06&       30.81&      30.75 &      220.2 \\ 
AEGIS-1 &      9875&      14      20       14.32&+52      59       42.61&2008-06-23 22:54:14&       25.21&      25.20 &      231.2 \\ 
AEGIS-1 &      9876&      14      20       14.21&+52      59       43.03&2008-06-22 00:22:03&       33.29&      33.28 &      229.2 \\ 
AEGIS-2 &      9454&      14      19       14.72&+52      48       22.75&2008-09-11 04:47:10&       59.81&      56.80 &      310.7 \\ 
AEGIS-2 &      9455&      14      19       14.72&+52      48       22.78&2008-09-13 19:38:46&       100.20&     99.72 &      310.7 \\ 
AEGIS-2 &      9456&      14      19       15.06&+52      48       29.45&2008-09-24 08:15:30&       58.82&      58.35 &      325.2 \\ 
AEGIS-2 &      9457&      14      19       11.06&+52      48       10.35&2008-06-27 07:08:38&       32.75&      32.74 &      235.7 \\ 
 AEGIS-2 &     9727&      14      19       14.72&+52      48       22.82&2008-09-12 16:44:12&       36.16&      34.94 &      310.7 \\ 
AEGIS-2 &      9729&      14      19       11.31&+52      48        09.80&2008-07-09 16:47:58&       48.29&      48.04&      240.2 \\ 
AEGIS-2 &      9730&      14      19       15.06&+52      48       29.41&2008-09-25 16:50:54&       53.97&      53.72 &      325.2 \\ 
AEGIS-2 &      9731&      14      19       11.25&+52      48        09.91&2008-07-03 10:58:47&       21.38&      21.38&      239.2 \\ 
AEGIS-2 &      9733&      14      19       15.06&+52      48       29.44&2008-09-27 01:15:33&       58.82&      58.36 &      325.2 \\ 
 AEGIS-2 &     9878&      14      19       11.07&+52      48       10.42&2008-06-28 06:03:20&       15.74&      15.73 &      235.7 \\ 
AEGIS-2 &      9879&      14      19       11.07&+52      48       10.36&2008-06-29 03:39:20&       27.02&      26.80 &      235.7 \\ 
AEGIS-2 &      9880&      14      19       11.25&+52      48        09.87&2008-07-05 17:00:17&       29.89&      29.45&      239.2 \\ 
AEGIS-2 &      9881&      14      19       15.06&+52      48       29.47&2008-09-28 08:15:12&       54.53&      53.93 &      325.2 \\ 
AEGIS-3 &      9458&      14      18        06.09&+52      37       17.03&2009-03-18 12:20:16&        6.66&      6.65 &      136.9 \\ 
AEGIS-3 &      9459&      14      18       12.11&+52      36       57.84&2008-09-30 19:20:28&       69.91&      69.55 &      329.7 \\ 
AEGIS-3 &      9460&      14      18       12.12&+52      36       58.08&2008-10-10 06:17:49&       21.91&      21.36 &      330.2 \\ 
AEGIS-3 &      9461&      14      18        07.78&+52      36       37.50&2009-06-26 09:30:12&       23.73&     23.73 &      230.2 \\ 
AEGIS-3 &      9734&      14      18       11.83&+52      36       50.93&2008-09-16 11:01:21&       49.98&      49.47 &      315.2 \\ 
AEGIS-3 &      9735&      14      18       11.83&+52      36       50.95&2008-09-19 03:14:15&       50.00&      49.47 &      315.2 \\ 
AEGIS-3 &      9736&      14      18       11.83&+52      36       50.97&2008-09-20 11:07:10&       50.13&      49.48 &      315.2 \\ 
AEGIS-3 &      9737&      14      18       11.83&+52      36       50.94&2008-09-21 17:53:00&       49.99&      49.48 &      315.2 \\ 
AEGIS-3 &      9738&      14      18       12.11&+52      36       57.84&2008-10-02 06:56:22&       61.89&      60.60 &      329.7 \\ 
AEGIS-3 &      9739&      14      18       12.12&+52      36       58.09&2008-10-05 11:28:12&       42.91&      42.59 &      330.2 \\ 
AEGIS-3 &      9740&      14      18        06.30&+52      37       19.96&2009-03-09 22:24:18&       20.38&     20.37 &      130.2 \\ 
AEGIS-3 &     10769&      14      18        05.99&+52      37       14.25&2009-03-20 13:38:26&       26.69&     26.68 &      143.0 \\ 
AEGIS-3 &     10847&      14      18        09.85&+52      37       32.39&2008-12-31 05:06:27&       19.27&     19.27 &       57.2 \\ 
 AEGIS-3 &    10848&      14      18        09.86&+52      37       32.43&2009-01-01 17:11:57&       17.91&     17.91 &       57.2 \\ 
AEGIS-3 &     10849&      14      18        09.85&+52      37       32.52&2009-01-02 21:25:57&       15.93&     15.92 &       57.2 \\ 
AEGIS-3 &     10876&      14      18        06.30&+52      37       19.99&2009-03-11 01:37:20&       17.21&     17.21 &      130.2 \\ 
AEGIS-3 &     10877&      14      18        06.31&+52      37       20.03&2009-03-12 15:15:57&       16.23&     16.22 &      130.2 \\ 
AEGIS-3 &     10896&      14      18       07.50&+52      36       38.72&2009-06-15 18:46:14&       23.29&      23.29 &      224.7 \\ 
AEGIS-3 &     10923&      14      18       07.77&+52      36       37.40&2009-06-22 07:38:22&       11.62&      11.62  &      230.2 \\

\enddata
\tablenotetext{a}{Field Name: note that there is an approximate correspondence between, respectively AEGIS1-3 and EGS 3-5, in the AEGIS-XW survey of L09}
\tablenotetext{b}{\chandra\ Observation ID}
\tablenotetext{c}{Nominal position of pointing (J2000)}
\tablenotetext{d}{Start date and time (UT)}
\tablenotetext{e}{Raw exposure time}
\tablenotetext{f}{Exposure time after data screening as described in \S\ref{sec:reduction}}
\tablenotetext{g}{Roll angle in deg of the observation}
\end{deluxetable}

\section{Observations and Data Reduction}
\label{sec:data_reduction}

\subsection{X-ray data}

The new AEGIS-XD  \chandra\ data were taken at three nominal pointing positions, which we have designated AEGIS-1, AEGIS-2 and AEGIS-3. Each field was approved to receive approximately 600ks in exposure as part of Chandra Cycle 9, supplementing the $\sim 200$ks exposures acquired in Cycle 3 (Nandra et~al. 2005) and Cycle 6 (L09).  The new exposures were split up into smaller segments  ranging in duration from $\sim 7-100$~ks. These observations were all taken in the time period 2007 Dec 11 to 2009 June 26  using the ACIS-I instrument (Garmire et~al. 2003) without any grating in place. All new AEGIS-XD data were taken in VFAINT mode. Full details of the new \chandra\ observations are given in Table~\ref{tab:obs}. The three subfields AEGIS-1, 2, and 3 were defined as the spatial limits of ObsIDs 9450, 9454 and 9458 respectively, with a border of $\pm 20$ pixels in the X and Y directions. These sub-fields were analyzed separately but have considerable overlap: common sources were removed at a later stage. 

The total exposure times for the Cycle 9 pointings before screening were 583ks (AEGIS-1),  597ks (AEGIS-2) and  596ks (AEGIS-3). The centre of these fields correspond fairly closely to those of the EGS-3, EGS-4 and EGS-5 fields of L09. Here we add together the data from the 200ks EGS 3-5 fields together with our new data, to a nominal 800ks depth in each field. 
For the AEGIS-1 sub-field there is some overlap with the EGS2 data of L09, but this is very small and at large off-axis angles, so these data were not included. For the AEGIS-3 sub-field, we also add together 3 ObsIDs from the original ``Groth-Westphal Strip'' (GWS) survey of Nandra et~al. (2005), which were taken in FAINT mode. These have significantly different pointing centers and roll angles to the L09 AEGIS-XW survey, but there is substantial overlap of the tiles with the AEGIS-3 field at one edge. A list of the ObsIDs overlapped with each field is given in Table~\ref{tab:egs_obs}. Note that the final catalogue presented here consists only of the sources in the region covered by the new X-ray imaging (i.e. covered by the ObsIDs listed in Table~\ref{tab:obs}), otherwise they should be present in the catalogue of L09. 

The total (pre-screening) exposure times for each field were then 779ks (AEGIS-1), 787 ks (AEGIS-2) and 782ks (AEGIS-3, excluding the GWS/EGS-8 exposure which only covers a relatively small part of the field). The data and analysis of the 200ks imaging of the AEGIS-XW is fully described in L09. Here we largely follow the same reduction analysis methodology described in that paper, which we now describe in brief, noting any changes. 


\begin{deluxetable}{cc}
\tabletypesize{\small}
\tablewidth{0pt}
\tablecaption{AEGIS-XW 200ks  ObsIDs combined with new AEGIS-XD fields (Table 1) \label{tab:egs_obs}}
\tablehead{
\colhead{AEGIS-XD field} & 
\colhead{AEGIS-XW 200ks ObsIDs} 
}
\startdata
AEGIS-1 & 5845, 5846, 6214, 6215 \\
AEGIS-2 & 5847, 5848, 6216, 6217 \\
AEGIS-3  & 3305, 4357, 4365, 5849, 5850, 6218, 6219
\enddata
\end{deluxetable}







\subsection{X-ray data reduction}
\label{sec:reduction}

The data reduction was performed using the CIAO data analysis software v4.1.2 (Fruscione et~al. 2009) and follows the scheme described in L09, with some minor changes and improvements, as detailed here. Briefly, for each individual Obsid, we corrected the data for aspect offsets, applied the bad pixel removal and destreaking algorithms, removed cosmic ray afterglows using standard tools, and corrected for CTI and gain effects. We also applied the ACIS particle background cleaning algorithm to the VFAINT mode data. Analysis was restricted to ACIS chips 0-3, and ASCA-style event grades 0,2,4 and 6. To reject periods of high background we used the procedure of Nandra et~al. (2007), adopting a threshold of 1.4 times the quiescent background level, determined as the count rate at which the background shows zero excess variance over that expected from statistical fluctuations alone. As in L09 and Nandra et~al. (2005) we relaxed this criterion in ObsID 4365 which contains a period of relatively high but stable background. As in L09, a flare was also manually removed from ObsID 5850. 

\begin{figure}[t]
\begin{center}
\epsscale{1.0}
\scalebox{0.35}
{\includegraphics{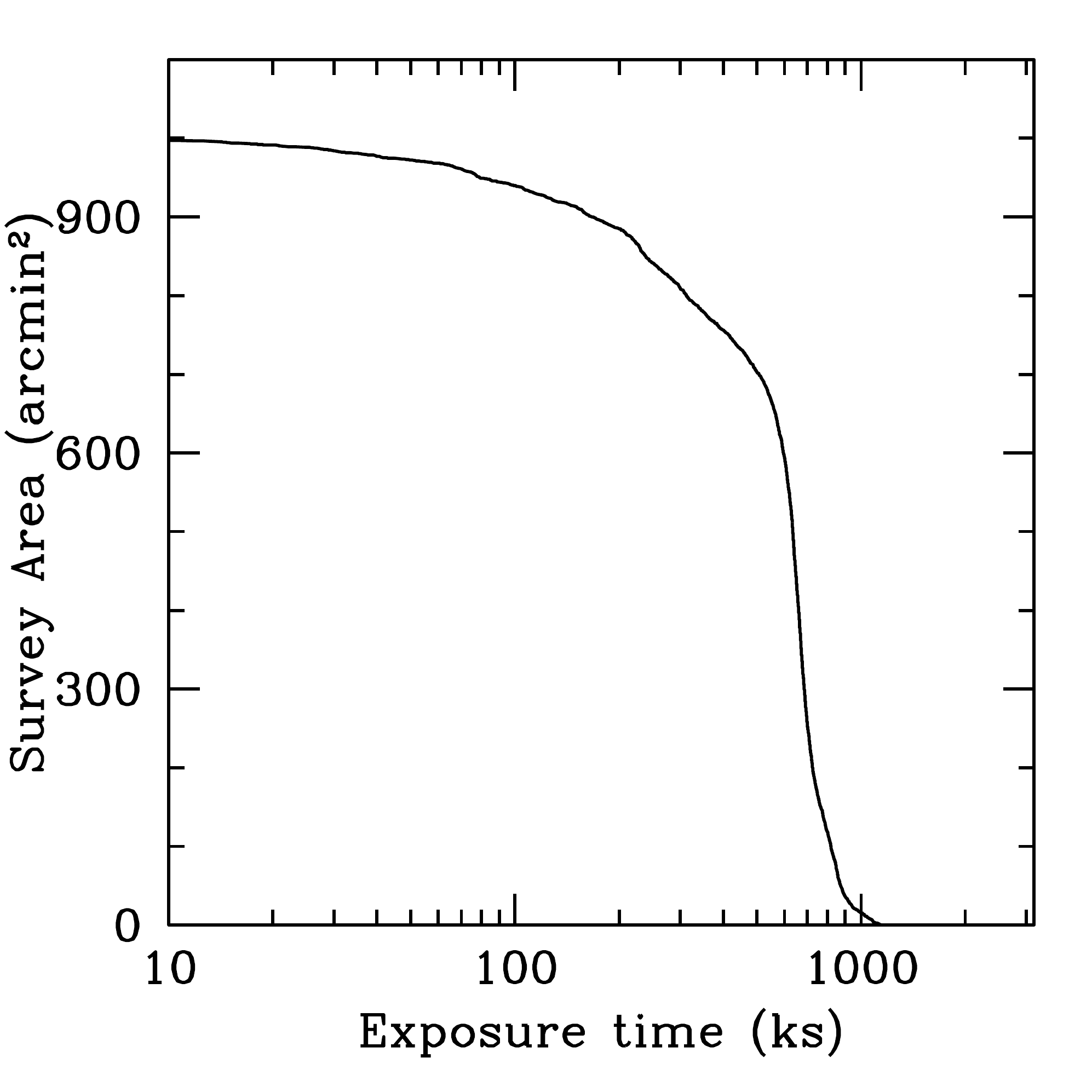}}
\caption{Effective exposure time as a function of survey area for the AEGIS-XD observations. 
\label{fig:exposure}}
\end{center}
\end{figure}

Following this basic reduction and screening, the astrometry of the individual image frames was corrected using a reference catalogue. Specifically, we use the CFHTLS i-band selected catalogue to register the AEGIS-XD images to the optical reference frame. We first ran the {\chandra} wavelet source detection task {\tt wavdetect} on the 0.5-7 keV image, using a detection threshold of $10^{-6}$, then used the CIAO task {\tt reproject\_aspect} to correct the astrometry compared to the reference image, and create new aspect solution files. The new aspect solutions were then applied to the original event files. The parameters used for {\tt reproject\_aspect} were a source match radius of 12Ó and a residual limit of 0.5Ó. Typically around 100 sources were detected in the individual ObsIDs with around 60~per cent of these having a counterpart used in the reprojection. The absolute value of the offset was typically small ($<0.5"$). Following this step we created event files for the individual frames in the standard full (FB; 0.5-7 keV), soft (SB; 0.5-2 keV), hard (HB; 2-7 keV) and ultrahard (UB; 4-7 keV) bands, and exposure maps for each were produced for each energy band using the {\tt merge\_all} task. The exposure maps were created at multiple energies with weights appropriate for a $\Gamma=1.4$ power law spectrum (see Table~\ref{tab:exp_weights}). The individual ObsID images and exposure maps were then stacked together. At the native resolution of ACIS (0.492" pixels), images of the entire AEGIS-XD region would be too large to manipulate efficiently. We therefore created separate image stacks for the AEGIS-1, AEGIS-2 and AEGIS-3 fields. These overlap significantly, meaning edge effects can be avoided, but in the final catalogue we remove duplicated entries detected in more than one image (see below). Exposure maps for the stacked images were created by summing the exposure maps of the individual ObsIDs. Figure \ref{fig:exposure} shows the effective exposure time as a function of survey area for the AEGIS-XD data.

\begin{deluxetable}{lllll}
\tabletypesize{\small}
\tablewidth{0pt}
\tablecaption{Weights used for exposure map calculations\label{tab:exp_weights}}
\tablehead{
\colhead{Energy} & 
\multicolumn{4}{c}{Energy band} \\
\colhead{(keV)} & 
\colhead{Full} & 
\colhead{Soft} & 
\colhead{Hard} &
\colhead{Ultrahard}
}
\startdata
0.65  & 0.2480 & 0.3867 & \ldots & \ldots \\ 
0.95   & 0.1509 & 0.2352 & \ldots & \ldots \\
1.25  & 0.1042 & 0.1625 & \ldots & \ldots \\
1.55  & 0.0776 & 0.1209 & \ldots & \ldots \\
1.85  & 0.0607 & 0.0947 & \ldots & \ldots \\
2.50  & 0.1359 & \ldots & 0.3789 & \ldots \\
3.50  & 0.0842 & \ldots & 0.2346 & \ldots \\
4.50  & 0.0590 & \ldots & 0.1645 & 0.4256 \\
5.50  & 0.0445 & \ldots & 0.1240 & 0.3208 \\
6.50  & 0.0352 & \ldots & 0.0980 & 0.2536 \\
\enddata
\end{deluxetable}

\begin{figure}[t]
\begin{center}
\epsscale{1.0}
\scalebox{0.5}
{\includegraphics{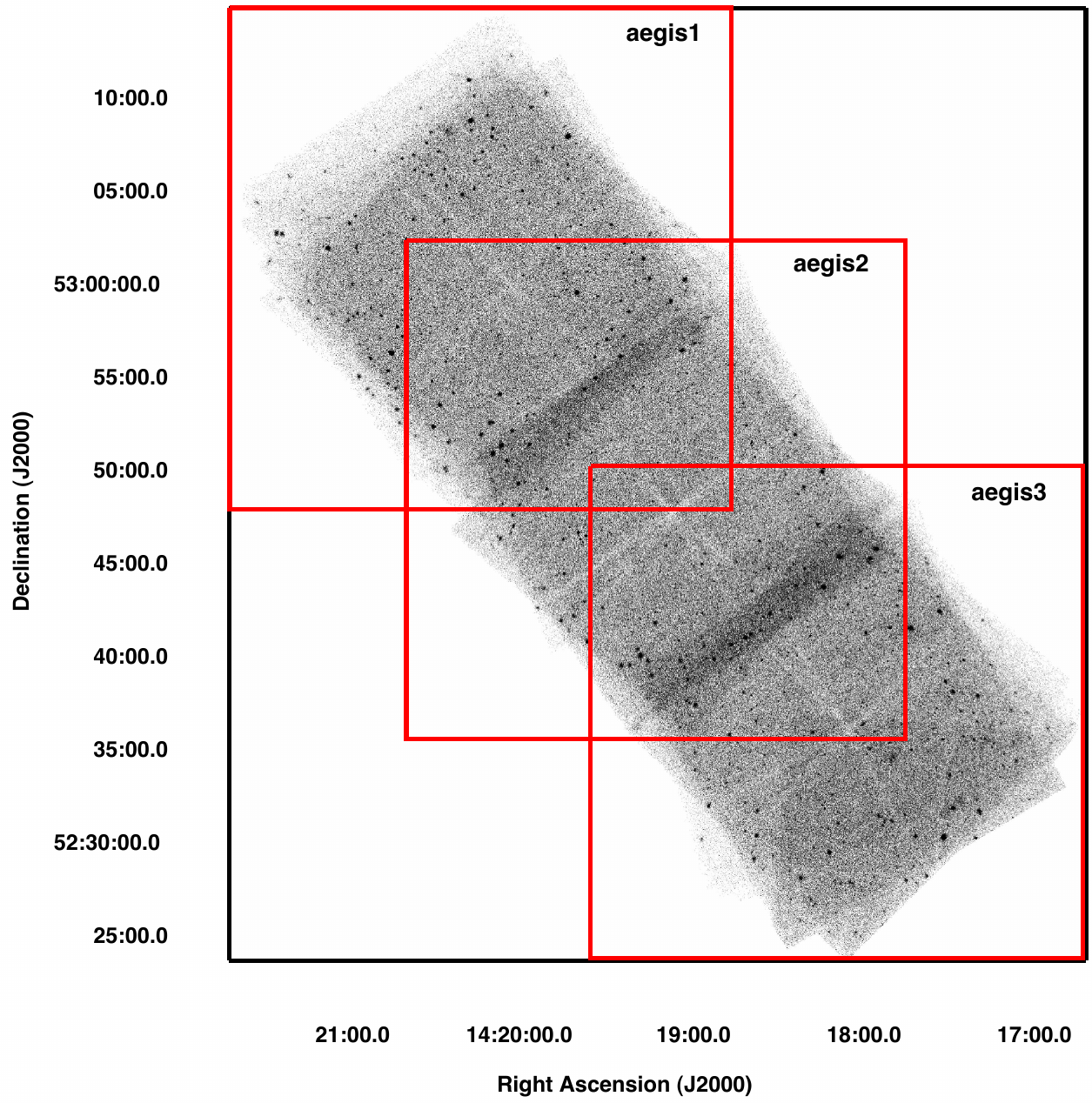}}

\caption{Mosaic full band image of the Extended Groth Strip showing the location and overlap of the 3 central AEGIS-XD fields, which have nominal 800ks depth. The sub-fields AEGIS-1, AEGIS-2 and AEGIS-3 are shown as red squares.   
\label{fig:image}}
\end{center}
\end{figure}

\subsection{Source Detection}
\label{sec:source_detection}
Source detection proceeded the same fashion as that described in L09 and the reader is referred to that paper for full details. Briefly, a``seed" source catalogue was first created using the {\tt wavdetect} task run at a low probability threshold ($10^{-4}$) on the mosaic images. This low threshold is intended to capture all potential sources, but likely contains many spurious sources. Aside from this thresholding, the only information used from {\tt wavdetect} in the final catalogue is the source position. Using these positions for the candidate sources, counts were extracted from the mosaic images using a circular aperture with radius equal to the exposure-weighted 70\% encircled energy fraction (EEF) of the {\it Chandra} point spread function (PSF). The PSFs were taken from a lookup table calculated using the MARX simulation software as described in L09. Background was determined using an annulus with inner radius equal to 1.5 times the 95\% EEF at the source position and outer radius 100 pixels larger than this, excluding detected sources. The background was then scaled to the size of the source region and the Poisson false probability of observing the total counts given the expected background was calculated, masking out the 95\% EEF of candidate sources. A significance threshold of $4 \times 10^{-6}$ was then applied, and a further detection iteration performed masking out only sources more significant than this. This iteration assures that the background is not underestimated due to the masking of random positive variations identified as candidate sources in the {\tt wavdetect} seed list. Any source detected at this $4 \times 10^{-6}$ probability level in the second iteration in any individual band was included in the final catalogue. The sources considered significant were band-merged using the matching criteria specified in Table 2 of L09. Photometry was then performed to estimate the fluxes in several energy ranges, even if the source was not considered a significant detection in that particular band.  After performing the band merging we visually inspected the images of the sources in each of the fields and checked that the correct cross-band counterparts were identified. Two sources in the catalogue were detected in the soft and full bands but not correctly matched with their hard (and in one case ultra-hard) band counterparts. These were combined manually. Two single-band detected sources (one soft, one ultra hard) were removed from the catalogue after visual inspection revealed strong contamination by nearby bright sources. These removed sources were most likely incorrectly identified in the initial {\tt wavdetect} seed catalogue. 

\begin{deluxetable}{lccccc}
\tabletypesize{\small}
\tablewidth{0pt}
\tablecaption{Sources detected in one band but not another\label{tab:det_sources}}
\tablehead{
\colhead{Detection} & 
\colhead{Total number}&
\multicolumn{4}{c}{Non-detection band} \\
\colhead{band (keV)} & 
\colhead{of sources} &
\colhead{Full} & 
\colhead{Soft} & 
\colhead{Hard} &
\colhead{Ultrahard}
}
\startdata
Full (0.5--7)   & 859 & \ldots & 190    & 299    & 562 \\
Soft (0.5--2)   & 732 & 63     & \ldots & 282    & 478 \\
Hard (2--7 )    & 574 & 14     & 124    & \ldots & 277 \\
Ultrahard (4--7)& 299 & 2      & 45     & 2   &\ldots 
\enddata
\end{deluxetable}


Finally the source catalogs for the individual sub-fields AEGIS 1,2 and 3 were merged to remove duplicate sources in the overlapping regions. A 2" search radius was adopted, and as in L09 the source was chosen from the field with the smallest off-axis angle for the final catalog.

One significant change in the current paper compared to L09 is in the X-ray photometry. Specifically, here we have adopted elliptical apertures  to extract the counts from the individual ObsIDs, using the 90\%EEF PSF appropriate for the ObsID in question (95\% in the case of the soft band).  This contrasts with the source detection described above, and the photometry in L09, which adopt circular apertures. Fluxes were estimated using the Bayesian methodology described in L09, using a $\Gamma=1.4$ spectrum with Galactic $N_{\rm H}$ of  $1.3\times10^{20}$~cm$^{-2}$ (Dickey \& Lockman 1990). The count rates in the full, hard and ultrahard bands were also extrapolated to fluxes in standard energy  bands: 0.5--10, 2--10, and 5--10~keV, respectively. Hardness ratios were calculated using the Bayesian methodology BEHR (Park et~al. 2006).  

\subsection{Spitzer/IRAC observations}

The IRAC Guaranteed Time Observations of the AEGIS field cover a region approximately $2^{\circ}$ by $10^{\prime}$ (Barmby et~al. 2006, 2008). The ACIS field of view is wider than this (see Fig~\ref{fig:image}) meaning that the GTO IRAC observations miss the edges of the deep {\it Chandra} imaging. As shown by, e.g., L09, and discussed below, Spitzer IRAC observations are critical for secure identifications of X-ray sources. For this reason, as part of the {\it Chandra} program we obtained additional Spitzer IRAC imaging of the edges of the strip. The IRAC coverage map is shown in Fig.~\ref{fig:layout} and the data reduction was performed as described in Barro et~al. (2011), and is incorporated into the {\t Rainbow} database.



\section{Point Source Catalog}
\label{sec:results}

\begin{figure*}
\begin{center}
\epsscale{1.0}
\hspace{-0.5in}
\scalebox{0.2}
{\includegraphics{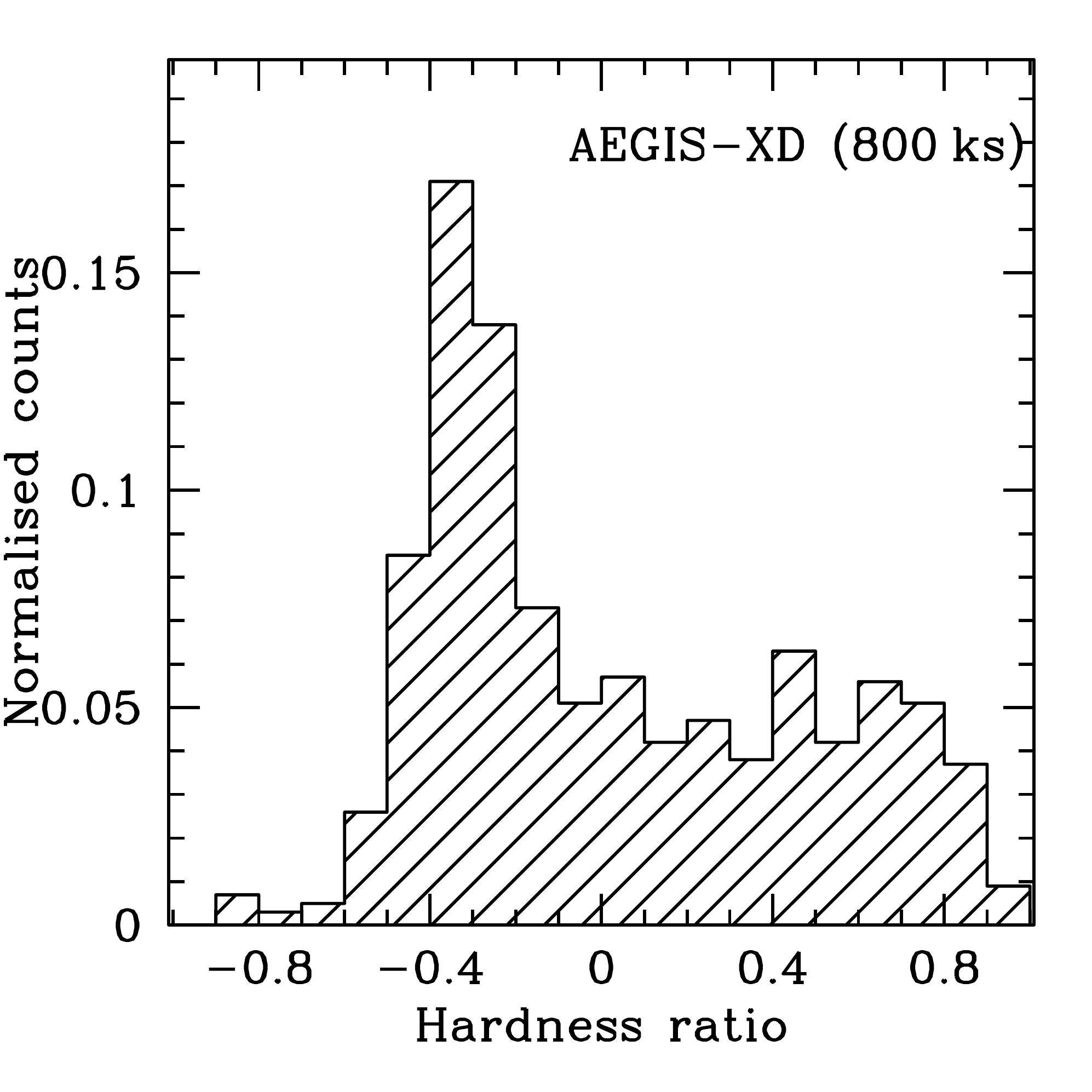}}
\scalebox{0.2}
{\includegraphics{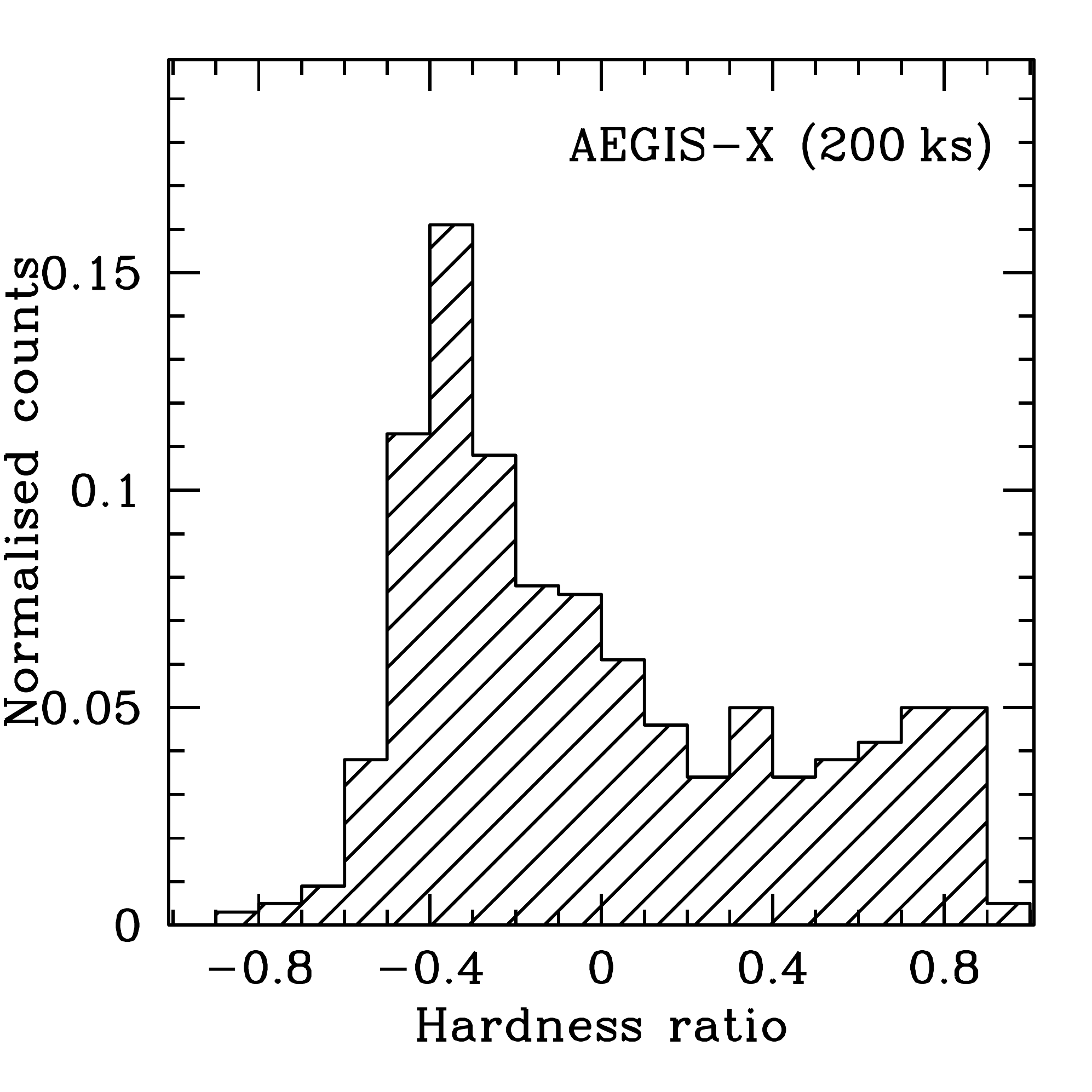}}
\scalebox{0.2}
{\includegraphics{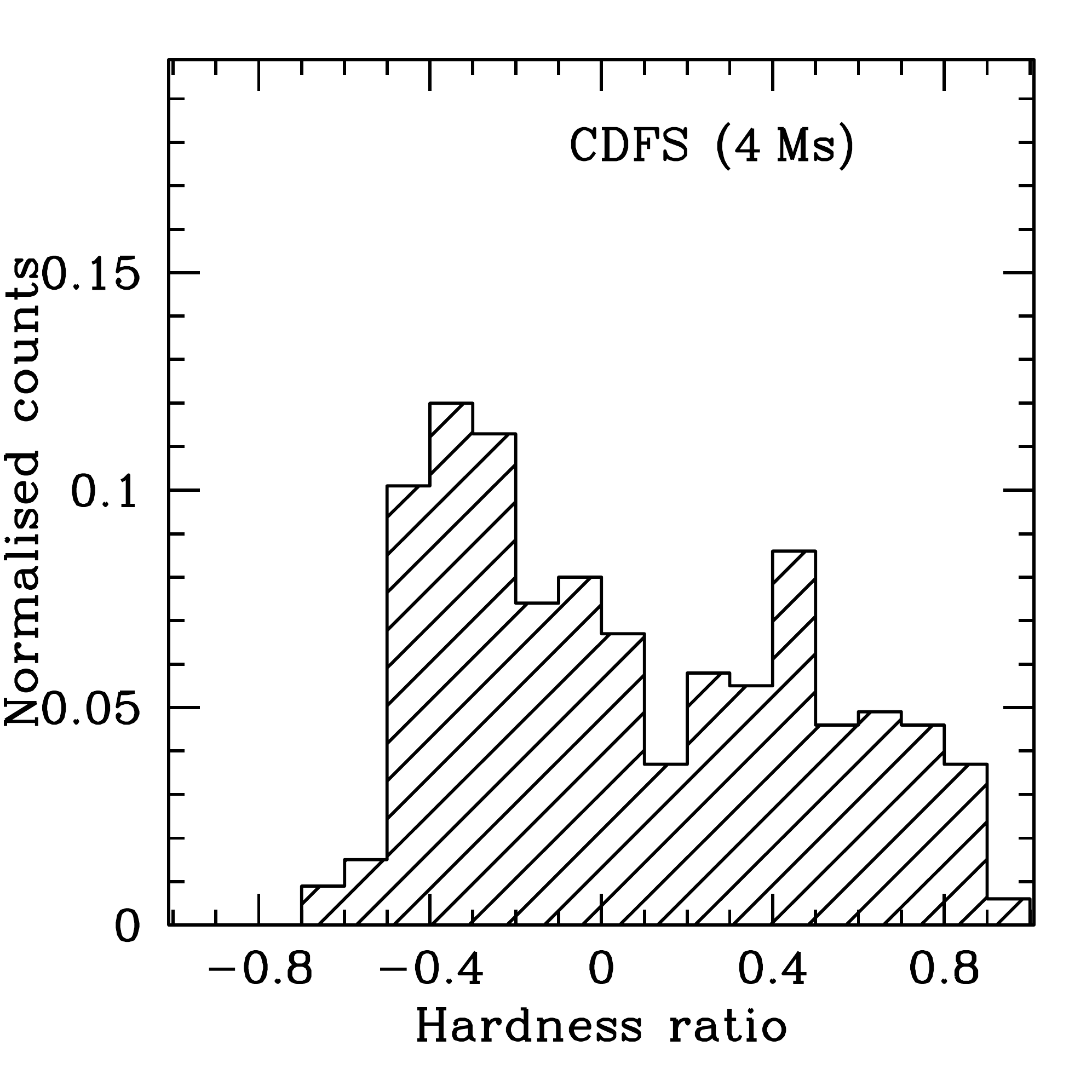}}

\caption{Hardness ratio distribution of hard-band (2-7\,keV) selected sources in the AEGIS-XD (left panel; this paper), AEGIS-X (middle panel; L09) and 4Ms CDFS (right panel; Rangel et~al. 2013). The hardness ratio is determined from the counts in the 0.5-2 and 2-7\,keV bands.  \label{fig:hist_hr}}
\end{center}
\end{figure*}

The final point source catalogue in the AEGIS-X Deep area consists of 937 sources. Of these, respectively 859, 732, 574 and 299 were detected at $p <4 \times 10^{-6}$ in the full, soft, hard and ultra hard bands. Sources detected in one band but not another are detailed in Table~\ref{tab:det_sources}. The AEGIS-XD X-ray source catalog with full X-ray information is made publicly available in FITS table format as detailed in an Appendix to this paper and at \url{http.mpe.mpg.de/XraySurveys/}. As a demonstration of the properties of the AEGIS-XD X-ray soruces, Figure \ref{fig:hist_hr} presents the hardness ratio distribution of the hard-band (2-7\,KeV) selected sample. This is compared to the hardness ratio sources of hard-band detected sources in the AEGIS-X (L09) and the 4Ms CDFS (Rangel et~al. 2013). Hardness ratios are estimated from the counts in the 0.5-2 and 2-7\,keV spectral bands.

\subsection{Sensitivity}

\begin{figure*}
\epsscale{1.0}
\hspace{-0.5in}
\begin{center}
\scalebox{0.3}
{\includegraphics{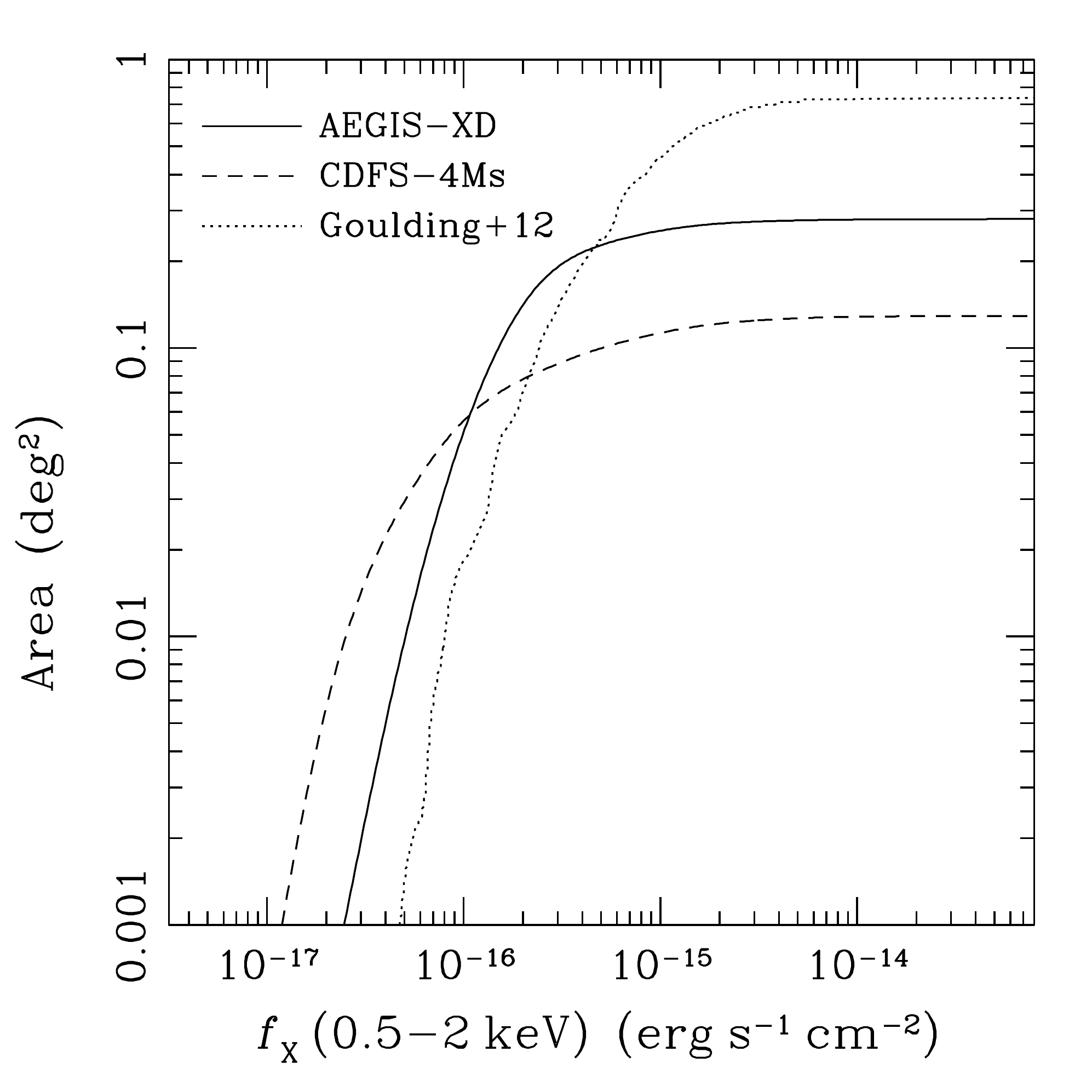}}
\scalebox{0.3}
{\includegraphics{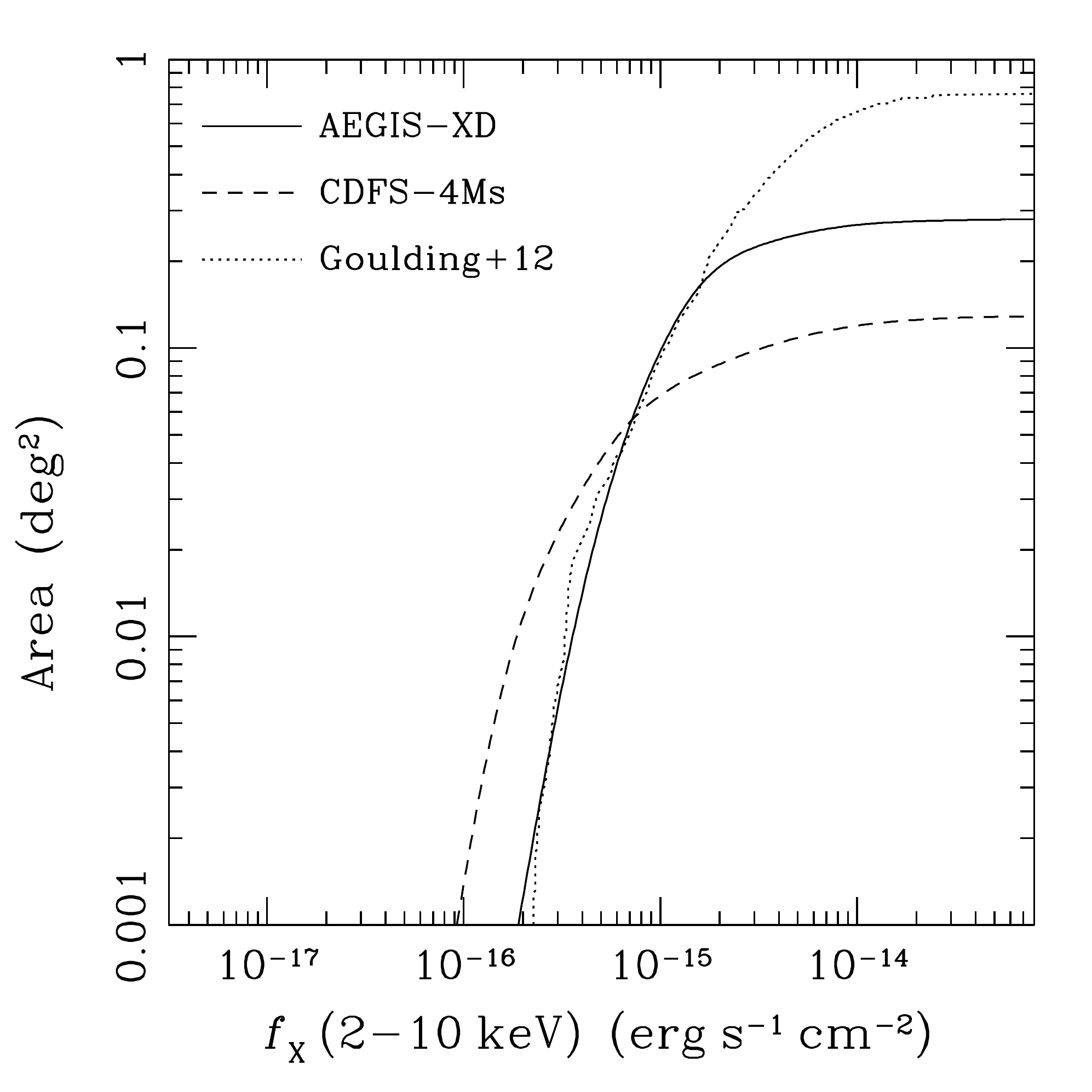}}
\scalebox{0.3}
{\includegraphics{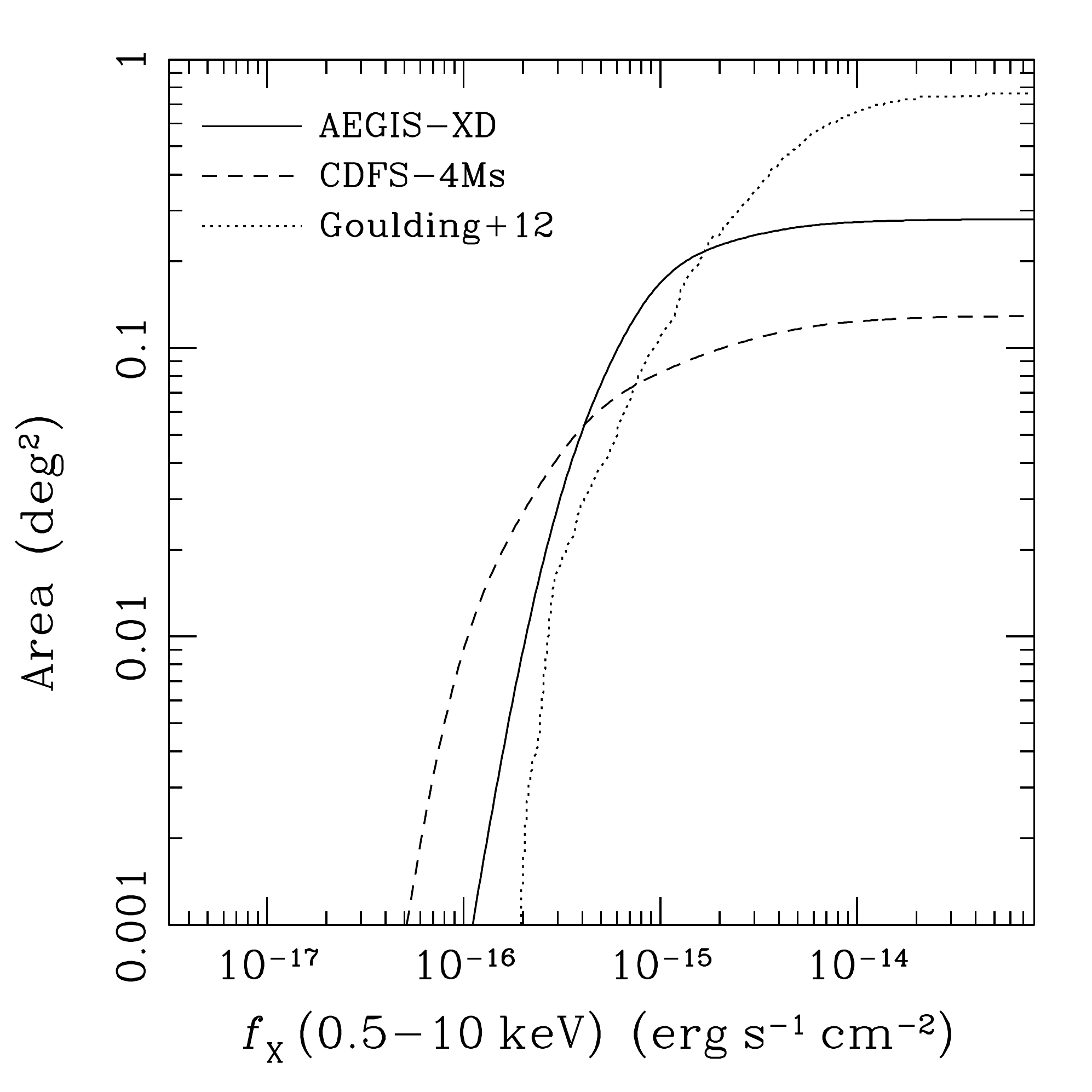}}
\scalebox{0.3}
{\includegraphics{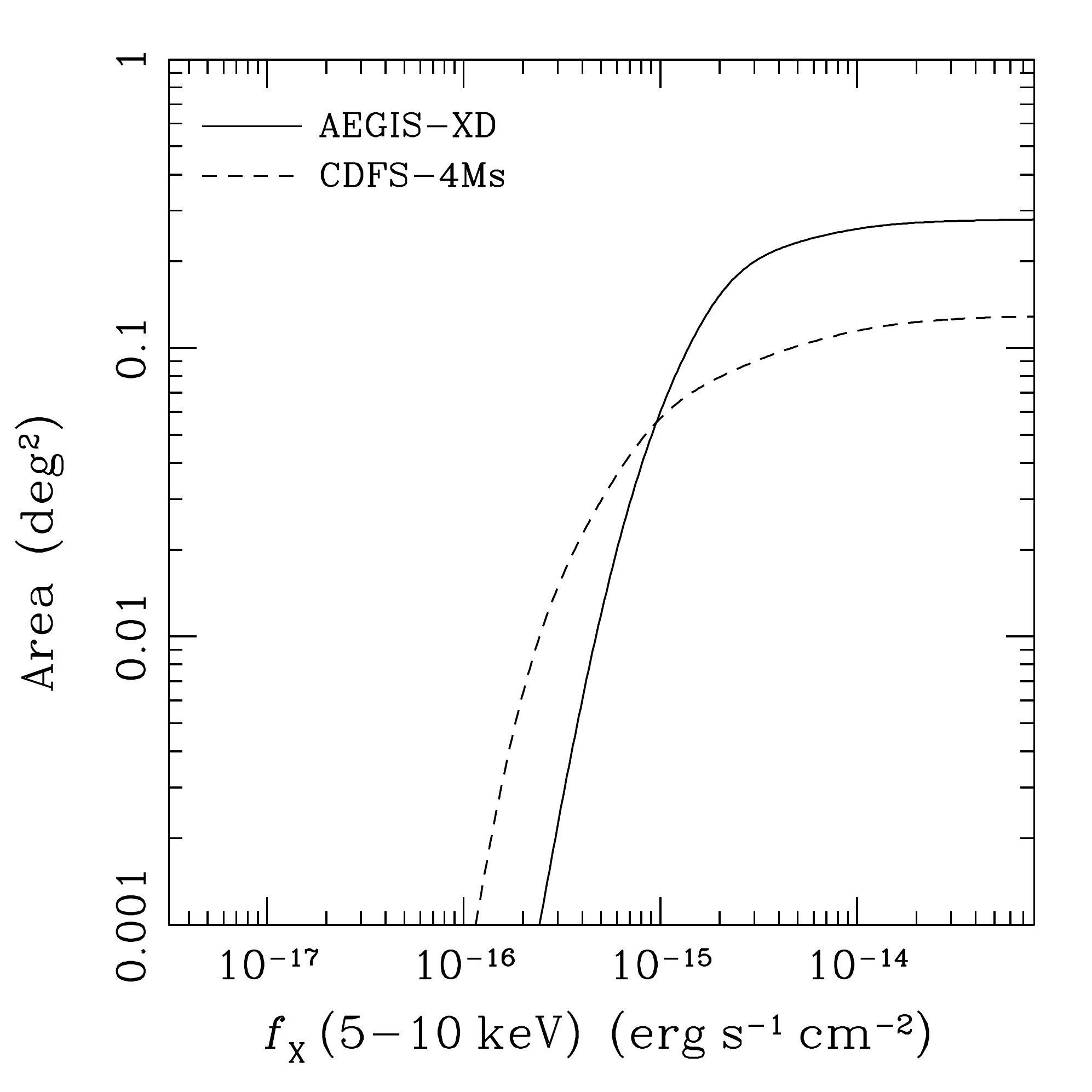}}

\caption{Sensitivity curves for the AEGIS-XD survey in the soft, hard, full and ultra hard bands (solid lines), calculated using the methodology of Georgakakis et~al. (2008b). These are compared to the 4Ms Chandra Deep Field South (dashed lines). The 4Ms CDFS reaches the deepest limits of any X-ray survey, but the AEGIS-XD data provide a considerable increase in area. Also shown in the plot are the sensitivity curves of the G12 in the entire Extended Groth Strip Chandra survey (deep and wide). For this comparison we apply conversions to account for the different power-law X-ray spectral index assumed by G12 to determine fluxes and the different energy bands used in their work compared to this  paper. Note, however, that their sensitivity calculation follows a different methodology as does their calculation of the exposure maps.
\label{fig:area_curve}}
\end{center}
\end{figure*}

Sensitivity maps were computed according to the procedure described in Georgakakis et~al. (2008b) as implemented by L09. This approach accounts for incompleteness and Eddington bias in the sensitivity calculation, which is performed in a manner which is also fully consistent with the source detection procedure. The flux limits for the new AEGIS-XD survey as a function of area are shown for various energy bands in Fig~\ref{fig:area_curve}, compared to the deepest X-ray survey in existence, the {\it Chandra} Deep Field South, together with the  sensitivity curves of the G12 source catalogue of the entire Extended Groth Strip Chandra survey. Following L09, we define the limiting flux of our observations as the flux to which at least 1\% of the survey area is sensitive. We find the limiting fluxes so defined to be $1.5\times10^{-16}$ (FB; 0.5--10~keV), $3.3\times10^{-17}$ (SB; 0.5--2~keV), $2.5\times10^{-16}$ (HB; 2--10~keV), and  $3.2\times10^{-16}$ erg~cm$^{-2}$~s$^{-1}$ (UB; 5--10~keV).  We also show in Table~\ref{tab:limits} the 50\% and 90\% completeness limits of the survey. 

\begin{deluxetable}{lccc}
\tabletypesize{\scriptsize}
\tablewidth{0pt}
\tablecaption{X-ray flux completeness limits for the AEGIS-XD survey\label{tab:limits}}
\tablehead{
&
\multicolumn{3}{c}{Completeness limit\tablenotemark{a}} \\
\colhead{Band}&
\colhead{1\%\tablenotemark{b}} &
\colhead{50\%\tablenotemark{b}} &  
\colhead{90\%\tablenotemark{b}}   
}
\startdata
Full &$1.46$ & $8.22$ & $35.9$\\
Soft &$0.33$ & $2.02$ & $ 9.22$\\
Hard &$2.48$ & $13.6$ & $58.2$ \\
Ultrahard & $3.27$ & $18.4$ & $82.2$ 
\enddata
\tablenotetext{a}{Flux to which 1, 50 and 90\% of the survey area is complete.}
\tablenotetext{b}{Fluxes are in units of $10^{-16}$ erg~cm$^{-2}$~s$^{-1}$.}
\end{deluxetable}

\begin{figure*}[t]
\epsscale{1.0}
{\includegraphics{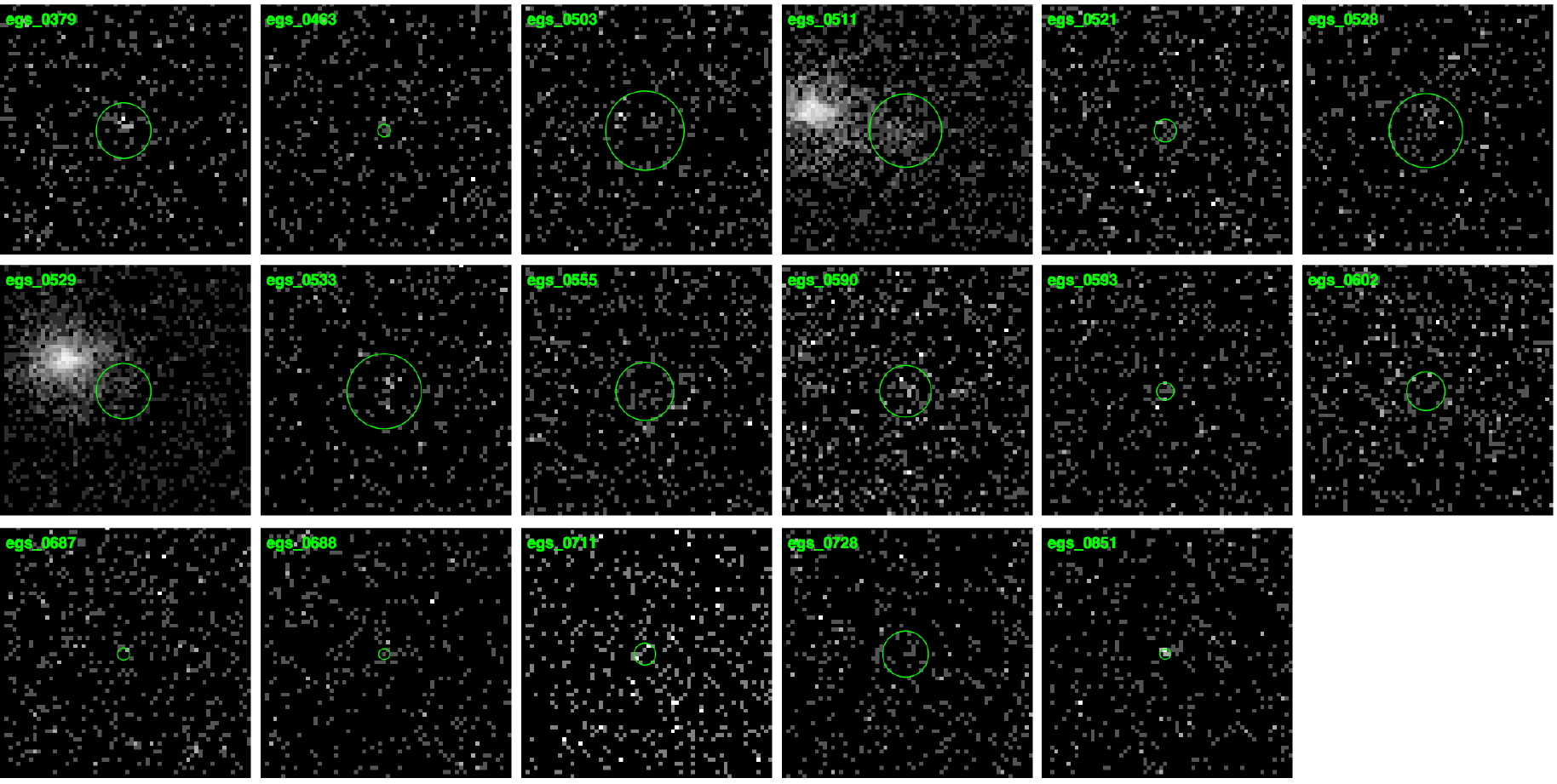}}
\caption{Sources in the 200ks AEGIS-XW catalogue of L09 not significantly detected in the deeper 800ks data. In each case the green circle shows the L09 position overlaid on the 800ks images from this work. Generally these are low significance detections which are not confirmed in the deeper data, but in a few cases the L09 detection appears in the wings of a nearby brighter source.  The size of the circle is equal to the 90\% EEF in the full band, while the cutouts have a size of 25x25".
\label{fig:l09_sources}}
\end{figure*}

\begin{deluxetable}{cccccccc}
\tabletypesize{\small}
\tablewidth{0pt}
\tablecaption{AEGIS-XW sources from L09 not included in AEGIS-XD catalog\label{tab:laird_nondet}}
\tablehead{
\colhead{L09} & 
\colhead{RA J2000\tablenotemark{a}} &
\colhead{Dec J2000\tablenotemark{a}} &
\colhead{$p_{\rm min}$\tablenotemark{a,b}} &
\colhead{Det.\tablenotemark{a}} &
\colhead{Bayesian flux\tablenotemark{a,c}} &
\colhead{$i\arcmin_{AB}$\tablenotemark{d}} &
\colhead{Classical flux limit\tablenotemark{e}}\\
\colhead{cat ID} & 
\colhead{(J2000)} & 
\colhead{(J2000)} & &
\colhead{bands} &
\colhead{($10^{-16}$~erg s$^{-1}$cm$^{-2}$)}&
\colhead{(mag)} &
\colhead{($10^{-16}$~erg s$^{-1}$cm$^{-2}$)}
}
\startdata
egs\_0379\tablenotemark{f} & 214.363649 & 52.525411 & $1.0\times10^{-8}$ & fs & $8.8^{+3.2}_{-2.7}$ &   25.12 & $<2.6$ \\
egs\_0463 & 214.491077 & 52.632192 & $3.6\times10^{-6}$ & s  & $<2.8$              &   18.43 & $<2.3$ \\
egs\_0503 & 214.530413 & 52.508574 & $3.3\times10^{-6}$ & s  & $<3.9$              &   21.78 & $<2.2$ \\
egs\_0511\tablenotemark{f} & 214.548133 & 52.759199 & $1.0\times10^{-8}$ & f  & $20.4^{+5.9}_{-5.3}$&   23.69 & $<3.7$ \\
egs\_0521 & 214.559843 & 52.568176 & $2.1\times10^{-6}$ & s  & $<8.2$              & $>27.0$ & $<4.2$ \\
egs\_0528 & 214.572498 & 52.446772 & $3.2\times10^{-6}$ & f  & $5.8^{+3.2}_{-4.2}$ &   23.79 & $<4.9$ \\
egs\_0529 & 214.573633 & 52.731312 & $1.0\times10^{-8}$ & fs & $9.2^{+3.9}_{-3.3}$ &   24.89 & $<2.6$ \\
egs\_0533 & 214.575732 & 52.442809 & $3.6\times10^{-6}$ & f  & $4.3^{+3.8}_{-2.8}$ &   22.00 & $<4.9$ \\
egs\_0555 & 214.597394 & 52.488107 & $1.6\times10^{-6}$ & s  & $<1.5$              &   21.97 & $<7.1$ \\
egs\_0590 & 214.643180 & 52.698915 & $4.8\times10^{-7}$ & fs & $4.7^{+3.1}_{-2.5}$ & $>27.0$ & $<1.8$ \\
egs\_0593 & 214.653435 & 52.627987 & $4.8\times10^{-7}$ & s  & $3.6^{+2.5}_{-1.9}$ &   22.23 & $<1.4$ \\
egs\_0602 & 214.662429 & 52.684361 & $3.3\times10^{-6}$ & s  & $<2.9$              &   25.39 & $<1.8$ \\
egs\_0687 & 214.813366 & 52.856652 & $3.0\times10^{-6}$ & h  & $2.4^{+1.9}_{-1.3}$ & $>27.0$ & $<1.4$ \\
egs\_0688 & 214.815105 & 52.793307 & $2.7\times10^{-6}$ & h  & $2.4^{+1.9}_{-1.7}$ & $>27.0$ & $<1.6$  \\
egs\_0711\tablenotemark{f} & 214.856439 & 52.745895 & $4.8\times10^{-7}$ & s  & $5.9^{+3.5}_{-3.3}$ &   24.28 & $<3.5$ \\
egs\_0728\tablenotemark{f} & 214.877813 & 53.007428 & $1.0\times10^{-8}$ & fs & $1.6^{+0.7}_{-0.7}$ &   21.54 & $<4.6$ \\
egs\_0851 \tablenotemark{f}& 215.076587 & 53.032611 & $2.6\times10^{-6}$ & s  & $3.2^{+2.5}_{-1.7}$ &   18.99 & $<1.4$ 
\enddata
\tablenotetext{a}{Values from L09.}
\tablenotetext{b}{Minimum false detection probability found for the four analysis bands.}
\tablenotetext{c}{Full band flux or upper limit.}
\tablenotetext{d}{Optical identification from L09 or upper limit where no counterpart exists}
\tablenotetext{e}{Values from this work.}
\tablenotetext{f}{Source also detected in G12.}
\end{deluxetable}

\subsection{False source estimation}

The source detection algorithm applied here (and in L09) is designed to provide an accurate estimate of the sensitivity of the X-ray observations at each position. The detection threshold can be altered depending on the number of likely spurious detections to be deemed acceptable in the catalogue. Here we adopt a relatively conservative threshold which should result in only a small number of spurious X-ray detections: L09, for example, estimated that $<1.5$~per cent of the AEGIS-XW 200ks sources were likely to be spurious. To assess this for the deeper AEGIS-XD data we performed tests of the source detection on simulated X-ray fields. The number of spurious sources expected in any given field may depend on the exposure map of the field: for example with multiple overlapping pointings, as is present here, edge effects might introduce problems into the source detection. 

The false source content of our catalogue was estimated based on the field configuration of the AEGIS-2 field, overlapping the individual subframes just as in the real data. The simulations were run initially just using the background, and then following the source procedure through as for the real observations. Correcting the area of the simulated observations to the total AEGIS-XD survey, we predict about 12 spurious sources in the catalogue, i.e. around 1.3\%, consistent with the estimates from L09.  We also estimated the false source contamination using simulated sources, based on the logN-logS function of Georgakakis et~al. (2008b), yielding similar results. 

A further estimate of the spurious source content can be made using the ultra-hard band (UB) images. Due to a combination of the strong energy-dependence of the effective area and PSF of the {\it Chandra} optics, inherent in the Wolter-1 design, and the nature of the spectra of the underlying source population, it is unlikely for real sources to be detected in only the UB, as it is the least sensitive of all the detection bands. 
As a result, UB-only sources must have heavy obscuration, at just the right level to suppress the soft, full and hard band detections below the threshold, but not so high that it suppresses the UB flux as well. Furthermore, in practice even heavily obscured sources are often detected in the soft X-ray band, due to the presence of scattered X-ray light (e.g. Brightman \& Nandra 2012). None of these considerations applies to spurious sources which, if present in the ultra-hard band, should not be detected in any other band. Just one source in our catalogue is detected only in the UB, and is thus a candidate for being spurious in the UB catalogue, which contains a total of 299 sources. This consideration suggests that the simulations might over-estimate the number of spurious sources in our catalogue. 

With our deeper X-ray data, we can also make a post-hoc check of the number of spurious sources in the 200ks L09 catalogue. Naively, it would be expected that all real sources in the AEGIS-XW catalogue also appear in our deeper observations in the overlapping area. This is not the case, however, if the sources are spurious, as the significance would tend to go down (and eventually below the detection threshold) with deeper data. There are in fact 17 sources in the AEGIS-XW catalogue, which are covered in the deeper {\it Chandra} pointings but are not listed as significant sources in our AEGIS-XD catalogue. These are listed in Table~\ref{tab:laird_nondet}, with postage-stamp images of the objects shown in Fig.~\ref{fig:l09_sources}. The images show that for the majority of the cases they are low significance detections in the 200ks catalogue which are not confirmed in the deeper data. Two (egs\_0511 and egs\_0529) are sources which were detected in the wings of nearby bright (and possibly extended) sources. A relatively large number of the others are in the very central regions of the image where the PSF is small, and where the original detections by L09 are based only on a very small number of photons. In these and indeed other cases the objects identified by L09 may not be truly spurious but simply represent sources whose true flux lies below the detection limit even of the 800ks data, but which due to Poisson statistics  generated a significant number of counts in the 200ks observation. In other words, they may be sources which were originally detected due to the Eddington bias. This assertion is supported by the fact that many of them have optical counterparts (see Table~\ref{tab:laird_nondet}. This may also be explained if the X-rays are variable, and the true flux has dropped by a large factor since the initial observations by L09. 

\subsection{Comparison with Goulding et~al. 2012}
\label{g12_xray}

G12 have already published a catalogue of X-ray sources and optical associations in this field. The set of {\it Chandra} data used is similar but not identical to ours, in that they analyse the entire AEGIS-X area, both deep and wide, while we use only the regions with nominal 800ks exposure. Further more we use two further ObsIDs in these areas, 9876 and 9881, not analysed by G12. The other 55 of our pointings are in common with G12, who nonetheless adopt a different procedure for data reduction, source detection and identification of the X-ray sources. A comparison of the latter can be found in Section~\ref{g12_opt}.  

\begin{figure}
\hspace{-0.05in}
\vspace{-0.5in}
\epsscale{1.0}
{\scalebox{0.3}
{\includegraphics{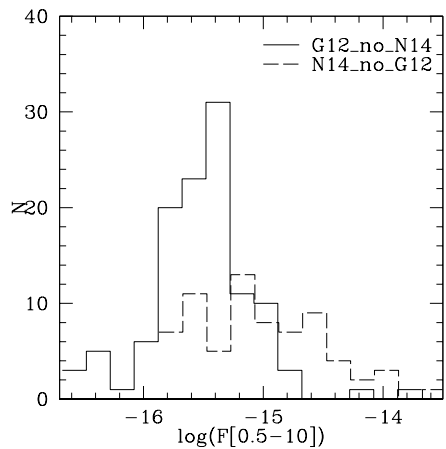}}
}
\caption{X-ray flux distribution of the sources that are in our catalog and not that of G12, or vice versa. G12 include a number of faint sources in their catalogue which do not satisfy our false probability threshold. A number of bright sources are included in our catalogue, but not in that of G12. These are generally at the edges of the field. where the PSF is relatively large. \label{fig:G12_N14}}
\end{figure}

We have cross-correlated the X-ray source list of G12 with our catalogue using a match radius of 3", and restricting to the common area. From this we find 115 sources in G12 that have no match in our catalogue. The vast majority (108) are faint source close to the detection threshold (see Fig~\ref{fig:G12_N14}). In this cases small differences in the analysis or source detection procedure (in particular the adopted detection threshold) can easily account for a detection in one catalogue, but not the other. In the photon-starved regime of {\it Chandra} the inclusion or exclusion of a single count in the source detection cell can easily change a ``detection" to a ``non-detection" or vice versa. Relatively minor differences in the determination of the background can have a similar effect. The other 7 sources in G12 but not in our catalogue are brighter and potentially a greater cause for concern. Visual examination of these shows three which are far off axis, and which do have a counterpart in our catalogue but just outside the 3" match radius. The remaining four are in crowded/bright source regions. Visual inspection suggest that these may indeed be distinct sources, but which have not been separated or identified as such by our detection algorithm. 

Performing the opposite comparison, we find 73 sources in our catalogue that have no counterpart within 3" in G12. An examination of Fig~\ref{fig:G12_N14} shows that a number of these sources are rather X-ray bright ($F_{\rm 0.5-10} > 10^{-15}$~erg cm$^{-2}$ s$^{-1}$). Examining the spatial distribution of these sources in our images, we find that the majority of them are at large off-axis angles, where the PSF is relatively broad.

\begin{figure}
\hspace{-0.05in}
\vspace{-0.5in}
\epsscale{1.0}
{\scalebox{0.3}
{\includegraphics{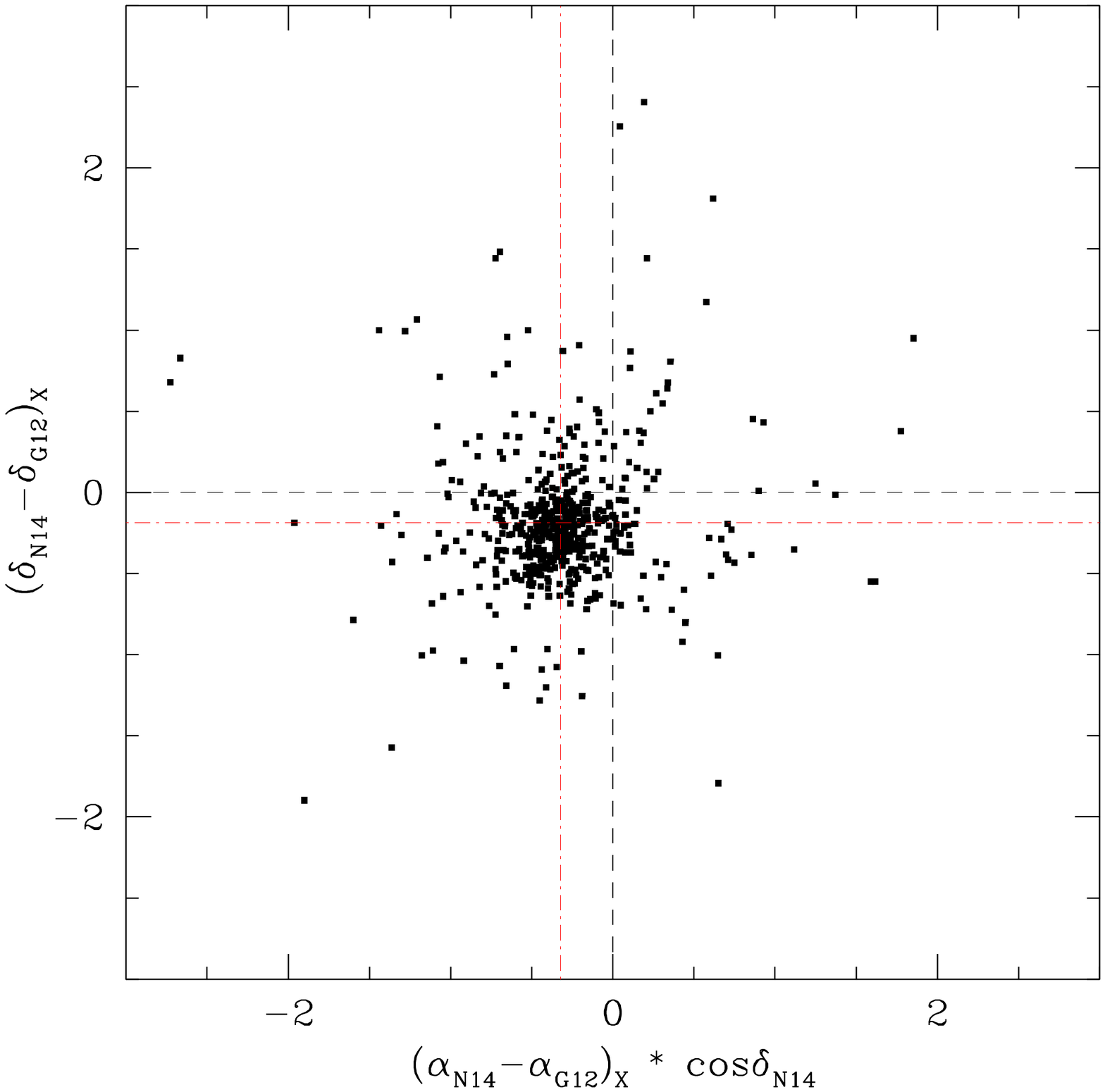}}
}
\caption{Offset between the X-ray positions of sources in our catalogue, and those in G12, in cases where the source is matched within 3" radius. There is a systematic offset of around 0.4". 
\label{fig:offset_x}}
\end{figure}

We have also compared the positions of our X-ray sources with those of G12, with the results of the comparison being shown in Fig.~\ref{fig:offset_x}. A systematic offset is found amounting to 0.37". This may be attributed in part to the different astrometric solutions adopted for the X-ray images, with our being tied to the CFHTLS i-band, and the G12 positions registered to the DEEP2 reference frame. 

\section{Multiwavelength counterparts and photometry}
\label{sec:rainbow}

We have identified multiwavelength counterparts to sources in our merged X-ray catalog using the likelihood ratio method 
(Ciliegi et~al. 2003; Brusa et~al. 2007; L09' Luo et~al. 2010) and photometry from the \textit{Rainbow} Cosmological Surveys Database\footnote{\url{http://rainbowx.fis.ucm.es/Rainbow\_Database/Home.html}} (Barro et~al. 2011a; Barro et~al. 2011b). The \textit{Rainbow} database is a compilation of the photometric datasets in several of the deepest extragalactic fields, including the AEGIS field. Table \ref{tab:photometry} lists the relevant datasets. The multiwavelength images were registered to a common astrometric reference frame and photometry was performed in consistent apertures to produce spectral energy distributions that span from the UV to mid-IR. 

To identify our X-ray sources, we first searched for counterparts in any of the multiwavelength images (based on SExtractor catalogs generated from each of the images) within 3.5" of the X-ray position. All the possible counterparts were then cross-matched to each other using a 2\arcsec\ search radius to create a single multiband catalog. Next, we performed photometry using the same elliptical (Kron) aperture across all the optical and near-infrared bands. The aperture was defined in a reference image for each source, typically the deepest available ground-based optical image. If the source had a Kron radius $<4.5$\arcsec\ then we extracted IRAC photometry using a 2\arcsec\ circular aperture and applied standard aperture corrections, thus accounting for the larger PSF of the IRAC images. If the source was detected in IRAC only, we applied a 1.5\arcsec\ aperture in the optical/near-IR images and forced a photometric measurement. If a single IRAC source was associated with multiple optical/near-IR counterparts, the positions of the optical/near-IR sources were adopted and used to deblend the IRAC photometry. The full procedure is described in Barro et~al. (2011a). 

We note that our procedure does not \emph{require} a significant detection in IRAC or any specific optical/near-IR band
and instead identifies potential counterparts to the X-ray sources in \emph{any} available image
(cf. the catalog presented by Barro et~al. 2011a, where an IRAC 3.6\micron\ or 4.5\micron\ detection was required).

\begin{deluxetable}{l r r c c r r}
\tabletypesize{\small}
\tablewidth{0pt}
\tablecaption{Multiwavelength photometric datasets included in the Rainbow database in our fields. \label{tab:photometry}}
\tablehead{
\colhead{Data}  & 
\colhead{$\lambda_\mathrm{eff}$ (\AA)} & 
\colhead{$Depth$ (mag$_{\rm AB}$)} & 
\colhead{$LR$ priority} &
\colhead{$N_\mathrm{cntrprt}$} &   
\colhead{A$\lambda$/E(B-V)} &
\colhead{zp$_{offset}$} \\
\colhead{(1)}  & 
\colhead{(2)} & 
\colhead{(3)} &
\colhead{(4)} &   
\colhead{(5)} &
\colhead{(6)}&
\colhead{(7)}
}
\startdata
IRAC [$3.6\mu$m]  & 35416.6& 23.9 & 1 &   881      (94.02\%)  & 0.162 & 0.086\\
IRAC [$4.5\mu$m]  & 44826.2& 23.9 & ... &  ...                               & 0.111  &0.114\\
IRAC [$5.8\mu$m]  & 56457.2& 22.3 & ... & ...                               &   0.076& 0.000 \\
IRAC [8\micron] &  78264.8& 22.3 &... &  ...                                   & 0.045 &0.000 \\
Subaru $R_c$ &    6486.4&26.1& 2 &   19  (2.03\%)                  & 2.549 & -0.022\\
ACS $V$ &    5796.7& 26.9 & ... &  ... &                                                2.852&-0.057\\ 
ACS $I$ &    8234.0& 26.1 &  3 &        0   (0.00\%) &                        1.825 & -0.020\\
Subaru $K_S$ &   21354.4& 23.7 &4 &  3   (0.32\%) & 0.367&0.055\\
CFHTLS $u^*$ &    3805.6& 25.7 &... &  ... & 4.675&0.066\\
CFHTLS $g'$ &    4833.7& 26.5 &... &  ... &3.615&-0.025 \\
CFHTLS $r'$ &    6234.1& 26.3 &... &  ... & 2.677&-0.018\\
CFHTLS $i'$ &    7659.1&25.9 & 5 &     3   (0.32\%)  &1.989& -0.012\\
CFHTLS $z'$ &    8820.9&24.7 & ... &  ... & 1.530&0.017\\
MMT $u'$ &    3604.1& 26.1 &... &  ... & 4.831&0.069\\
MMT $g'$ &    4763.5&26.7 & ... &  ... & 3.655&0.041\\
MMT $i'$ &    7770.5&25.3 & 6 &          1   (0.11\%) &1.932&-0.053 \\
MMT $z'$ &    9030.9& 25.3 &... &  ... & 1.473&0.101\\
DEEP $B$ &    4402.0&25.7 &... &  ... &4.096& -0.085\\
DEEP $R$ &    6595.1& 25.3& 7 &     19   (2.03\%) & 2.508&-0.012\\
DEEP $I$ &    8118.7&24.9& ... & ...  &  1.799 &-0.102\\
Palomar $J$ &   12435.0&21.9 & ... & ... & 0.893 & -0.012\\
Palomar $K_S$ &   21353.0&22.9& 8 &     3   (0.32\%) & 0.368&0.033\\
GALEX FUV &    1528.1& 25.6&... & ... &8.290& 0.000 \\
GALEX NUV &    2271.1& 25.6&... & ... & 8.612& 0.000\\
NICMOS F110W &   10622.4&23.5& ... & ... & 1.085& 0.000\\
NICMOS F160W &   15819.6&24.2& ... & ... &  0.593 &0.000\\
CAHA $J$ &   12029.7&22.9& ... & ... &  0.924 & 0.176\\
WFC3 $J$ &   12425.8& 27.4&... & ... & 0.875 & 0.050 \\
WFC3 $H$ &   15324.7& 27.5& ... & ... & 0.626 & 0.248 \\
NEWFIRM $J_1$ &   10441.4&25.1& ... & ...  & 1.164&-0.001\\
NEWFIRM $J_2$ &   11930.0&25.3& ... & ... & 0.940&0.024\\
NEWFIRM $J_3$ &   12764.0&24.5& ... & ... & 0.844&0.001 \\
NEWFIRM $H_1$ &   15585.4&24.1& ... & ... & 0.612& 0.019\\
NEWFIRM $H_2$ &   17048.8&24.4& ... & ... &  0.530&-0.003\\
NEWFIRM $K$      &   21643.9&24.2& ... & ... & 0.348&0.024\\
\enddata

\tablecomments{(1) Instrument and filter/band (2) Effective wavelength of the filter (3) Likelihood ratio priority. A higher priority means the counterpart is taken from that catalogue where a match is found in more than one (4) Number of counterparts assigned in that band (5) Galactic extinction (6) Zeropoint offset. The relevant references are listed in Barro et~al., (2011a and 2011b), except that for NEWFIRM data (Whitaker et~al. 2011) and CANDELS/ WFC3 (Grogin et~al. 2011, Koekemoer et~al. 2011).}
\end{deluxetable}

All of our X-ray sources (within the \textit{Rainbow} coverage) have at least one candidate counterpart within the 3.5\arcsec\ search radius, with $\sim 58$\% having 2 or more. 

In the next step we applied the likelihood ratio technique to determine which of these candidates are likely to be the true counterpart to the X-ray source, as opposed to a chance alignment. We first restricted the list of candidates to those with significant detections in a single, given optical, near-IR or mid-IR band with a measured magnitude, $m$. The likelihood ratio compares the probability that a candidate counterpart with magnitude $m$ found at a distance $r$ from the X-ray source position is the true counterpart and the probability that it is a spurious background source. 

The likelihood ratio ($LR$) is given by
\begin{equation}
LR= \frac{ q(m) f(r) } { n(m) }
\label{eq;lr}
\end{equation}
where $q(m)$ is the expected magnitude distribution of the true counterparts to the X-ray sources, $n(m)$ is the surface density of background sources as a function of magnitude, and $f(r)$ is the probability distribution of angular separations of the sources. We assume $f(r)$ can be described by a symmetric two-dimensional Gaussian distribution, 

\begin{equation}
f(r)=\frac{1}{2\pi\sigma^2}\exp\left(\frac{-r^2}{2\sigma^2}\right)
\label{eq:fr}
\end{equation}

where the standard deviation, $\sigma$, combined the X-ray and counterpart positional uncertainties, added in quadrature. The X-ray positional uncertainties used depend on the source counts and off-axis angle, as described in L09. As our \textit{Rainbow} counterpart catalog was limited to sources within 3.5\arcsec\ of an X-ray source, we estimated the background source density, $n(m)$, using the original SExtractor catalogs for the given band over the entire field, restricting to sources with significant detections and measured magnitudes, $m$, in that band. This approach provides an accurate estimate of the background source density using the entire photometric coverage.

\begin{deluxetable}{l c c c c c c c c c c c c c c}
\tablecaption{Summary of likelihood ratio matching results for AEGIS-XD \label{tab:matchstats}}
\tabletypesize{\scriptsize}
\tablewidth{0pt}
\tablehead{
\colhead{Catalog} & 
\colhead{Area/deg$^2$} & 
\colhead{$N_0$} & 
\colhead{$\sigma_0$} & 
\colhead{$L_\mathrm{th}$} & 
\colhead{$R$} & 
\colhead{$C$} & 
\colhead{$N_\mathrm{X}$} & 
\colhead{$N_\mathrm{ID}$} & 
\colhead{$N_\mathrm{NoID}$} & 
\colhead{$N_\mathrm{Multi}$} & 
\colhead{$N_\mathrm{Pri}$} \\
\colhead{(1)} &
\colhead{(2)} &
\colhead{(3)} &
\colhead{(4)} &
\colhead{(5)} &
\colhead{(6)} &
\colhead{(7)} &
\colhead{(8)} &
\colhead{(9)} &
\colhead{(10)} &
\colhead{(11)} &
\colhead{(12)} 
}
\startdata
RAINBOW [3.6\micron] &   0.26 &      65675 &   0.20 &   0.80 &   0.98 &   0.94 &    902 &    860 &     42 &    451 &    860\\
               Subaru $R$ &   0.28 &     126709 &   0.20 &   1.18 &   0.95 &   0.80 &    936 &    788 &    148 &    435 &     37\\
                  ACS $I$ &   0.14 &      61983 &   0.10 &   1.00 &   0.96 &   0.79 &    617 &    509 &    108 &    279 &      0\\
               Subaru $K$ &   0.08 &      10908 &   0.20 &   3.04 &   0.97 &   0.86 &    408 &    363 &     45 &    165 &      2\\
               CFHTLS $i$ &   0.26 &     116686 &   0.30 &   1.10 &   0.95 &   0.77 &    828 &    670 &    158 &    401 &      1\\
                  MMT $i$ &   0.29 &      93412 &   0.30 &   1.03 &   0.94 &   0.75 &    937 &    752 &    185 &    370 &      1\\
                 DEEP $R$ &   0.28 &      45968 &   0.30 &   4.06 &   0.97 &   0.45 &    922 &    430 &    492 &    312 &      0\\
            Palomar $K_s$ &   0.25 &      19086 &   0.30 &   1.58 &   0.96 &   0.76 &    881 &    693 &    188 &    202 &      6\\

\enddata
\tablecomments{Col 1:Catalog name and detection band used to match counterparts. Col 2: Area covered by both the multiwavelength data and deep X-ray data. Col 3: Number of multiwavelength sources in the X-ray area. Col 4:1 $\sigma$ positional accuracy of multiwavelength catalog in arcseconds. Col 5: Likelihood ratio threshold determined by the iterative procedure described in section \ref{sec:rainbow}. Col 6; Sample reliability, the mean of the individual reliabilities of each secure counterpart, as in Luo et~al (2010) Col 7: Sample completeness, the sum of the reliabilities for all counterparts divided by the total number of X-ray sources. Col 8: Total number of X-ray sources in the area covered by the multiwavelength data. Col 9: Number of secure counterparts to X-ray sources. Col 10: Number of X-ray sources without secure counterparts. Col 11: Number of X-ray sources with more than 1 candidate counterpart within the 3.5\arcsec\ search radius. Col 12: Number of X-ray sources assigned a primary counterpart in this band.}

\end{deluxetable}

The magnitude distribution of true counterparts, $q(m)$, was estimated via the iterative method described in Luo et~al. (2010). Briefly, a first estimate of $q(m)$ was determined by matching counterparts to X-ray sources within a small radius (we adopted 1.5\arcsec), and subtracting the background density. The $LR$ was then calculated for all counterparts to X-ray sources within our 3.5\arcsec\ search radius.  We found the counterpart with the highest $LR$ value for each X-ray source and applied a threshold, $LR_\mathrm{thresh}$, that maximises  the sum of the sample completeness and reliability to identify a sample of ``secure" matches. These secure matches were then used for a new estimate of $q(m)$, and the likelihood ratios recalculated. This process was repeated 10 times, resulting in a stable number of matches and threshold $LR$ values.

\begin{figure*}
\hspace{-0.050in}
\epsscale{1.0}
{\scalebox{0.8}
{\includegraphics{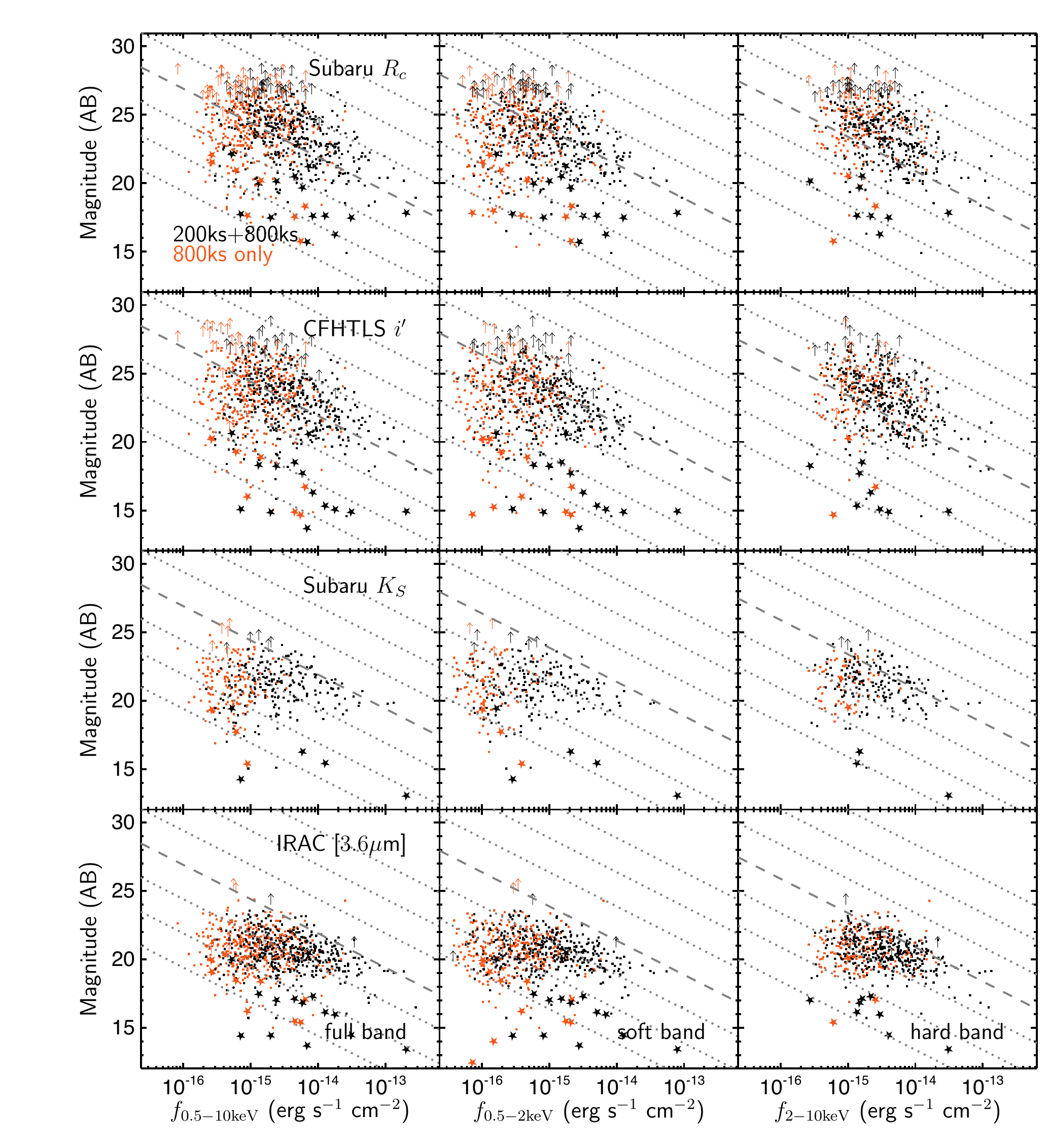}}
}
\caption{X-ray flux versus AB magnitude in various matching bands for all AEGIS-XD 800ks sources (orange symbols), and those also identified earlier in the shallower AEGIS -XW 200ks survey (black symbols). From top to bottom the comparison bands are the Subaru $R_{c}$-band, CFHTLS $i^{\prime}$, Subaru $K_{\rm s}$ and IRAC $3.6\mu$m. Anything with less than 3 sigma detection is shown as a limit (downward arrow) and stars are indicated with a star symbol. Note that the $K_{s}$ imaging covers a significantly smaller area than the other bands, accounting for the smaller bunker of matches sources. Lines denoting a constant ratio of the X-ray flux to the flux in the corresponding optical/IR band are shown as the dotted lines, with the dashed line representing a ratio of 1:1. 
\label{fig:fxfopt}}
\end{figure*}

We repeated the entire $LR$ matching process for the bands indicated in column 3 of Table \ref{tab:photometry}. Finally, we combined the matches to produce a master list of counterparts. We first took the secure counterparts in the highest priority band  (indicated by LR priority=1).  Next, we looped through the remaining match bands and assigned a final counterpart if a secure match was available in that band (and was not available in any of the higher priority bands).  No additional cross-matching is required to obtain the full multiband photometry as this is provided through matched apertures for all the \textit{Rainbow} counterparts, regardless of the detection band (c.f. Luo et~al. 2010). In practice, the vast majority ($\sim90$\%) of X-ray sources are assigned final counterparts from the highest priority match band, IRAC $3.6$\micron. IRAC is known to give the highest match rate for faint X-ray sources in deep {\it Chandra} surveys (Cardamone et~al. 2008; L09). The additional steps help us identify counterparts when the IRAC candidate is faint, blended, or non-existent. We assigned a match
to 929 of the 937 sources. 

The X-ray fluxes in the full, soft and hard bands are plotted against the counterpart magnitude in various matching bands in \ref{fig:fxfopt}. The lines of constant \fxfopt\ show a clear effect that the counterparts are generally brighter in the Subaru MOIRCS NIR ($K_{\rm s}$) and IRAC 3.6$\mu m$ bands than the are in the optical bands (CFHT $i^{\prime}$ or Subaru $R_{\rm c}$. This illustrates the well-known fact that the X-ray sources - which are dominated by AGN - reside in relatively red host galaxies (Barger et~al. 2003; Nandra et~al. 2007; Brusa et~al. 2009; Civano et~al. 2012), a fact which accounts for the much higher IRAC identification rate, compared to optical photometry. The figure also identifies objects which are newly detected in the 800ks survey, as opposed to those which were already in the 200ks catalogue of L09. While the fainter 800ks X-ray sources are typically identified also with fainter optical counterparts, a small but significant fraction of the faintest X-ray sources are identified with very bright optical sources (R$_{AB}$=20 or  brighter). Most of these sources are secure stars as classified via spectroscopy, multiple color-color selection (see Barro et~al. 2011a and 2011b) or SED fitting. Some of these sources are galaxies with a low X-ray Luminosity indicating that their X-ray emission may be dominated by stellar processes (e.g. via X-ray binaries and diffuse hot gas), rather than an accreting black hole.  Looking more specifically at this issue, we find a total of 49 sources with $F_{\rm X}/F{\rm opt}<-2$, where $F_{X}$ if the soft band flux and $F_{opt}$ is based on the Subaru $R$ band, which are not flagged as stars. 44 of these have a secure spectroscopic redshift ($z_{spec}>0$) while the other have reliable photo-z $>0$. All but one of these sources has $\log L_{\rm X}<42$ meaning that in principle they could be normal galaxies, rather than AGN. They represent $6.7$~per cent of the soft X-ray detected sample. This is consistent with the work of Lehmer et al. (2012), from which we would predict a normal galaxy fraction of 6.1~per cent at $F_{X}>1.96 \times 10^{-16}$~erg cm$^{-2}$ s$^{-1}$, our 1~per cent completeness soft flux limit. We nonetheless caution that separating AGN and normal galaxies based on these criteria is difficult, and requires detailed consideration of the properties of the individual objects.

There is also some evidence from this figure to suggest that the fainter X-ray sources do not become significantly fainter in the longer wavelength NIR or IRAC bands. The IRAC $3.6$\micron\ magnitude, for example, remains relatively constant over the full range of X-ray fluxes probed by the AEGIS-XD survey. The interpretation of this is not straightforward, but may be related to the fact that X-ray selection tends to identify the most massive galaxies at any given redshift (e.g. Bundy et~al. 2008; Aird et~al. 2013), and/or that the optical faintness of many of the X-ray sources is due to dust reddening.

\subsection{Astrometric accuracy}
\label{sec:astrom}

\begin{figure}
\begin{center}
\epsscale{1}
\scalebox{0.6}
{\includegraphics{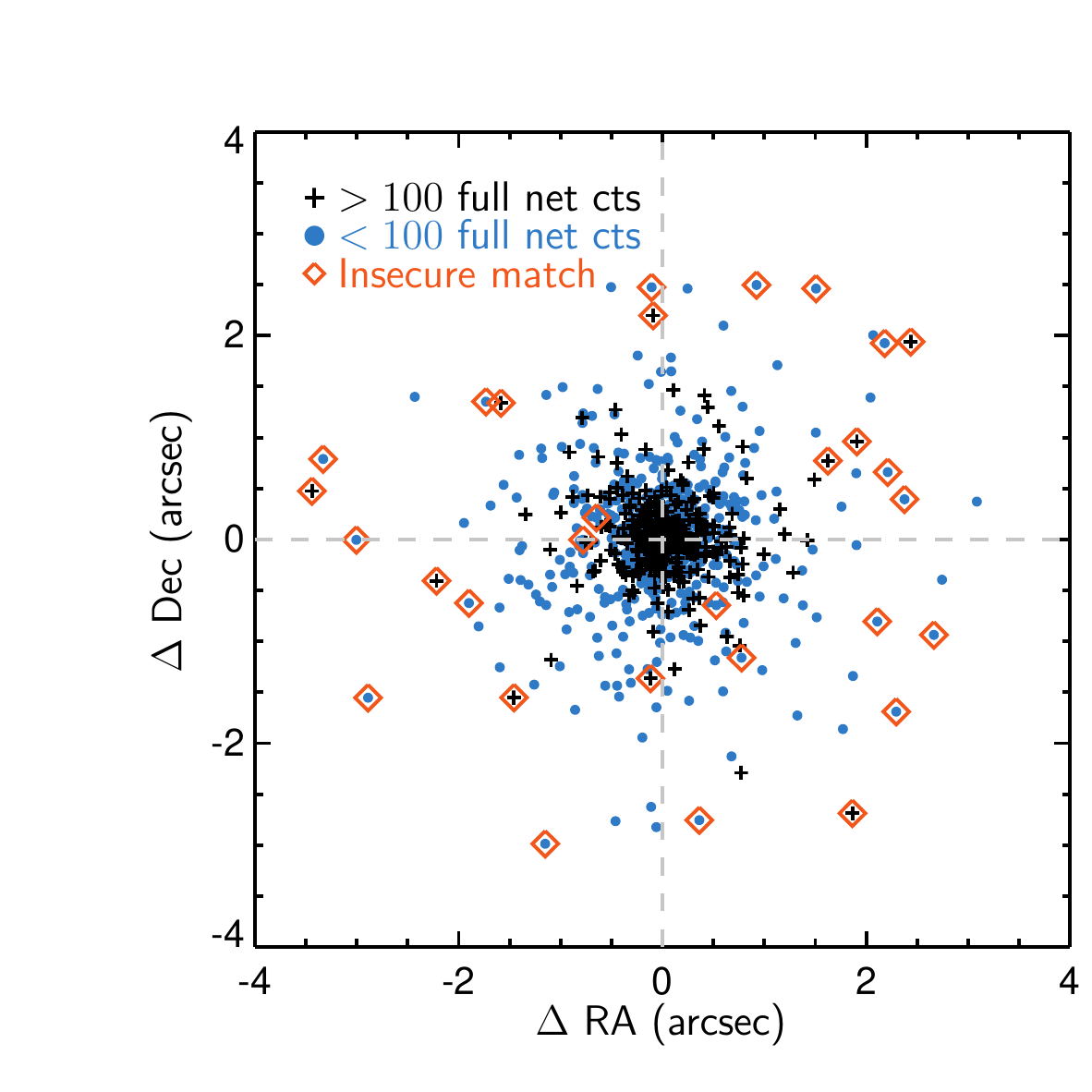}}

\caption{Offset between the X-ray and counterpart positions for the AEGIS-XD sources. Black crosses show bright objects where the X-ray position is statistically well determined. With $<100$ net counts the positional accuracies degrade somewhat. The vast majority of the secure counterparts ($>97$\%) none the less lie within 2" of the X-ray position.
\label{fig:astrom}}
\end{center}
\end{figure}

Following the cross-matching procedure we can make a post-hoc estimate of the astrometric accuracy of the X-ray positions in our catalogue. Fig.~\ref{fig:astrom} shows the offsets between the X-ray position and that of the multi band counterpart. Overall, we find that 84~per cent of the of the counterparts lie within 1" of the X-ray position, and 97~per cent within 2". As discussed by L09, the astrometric accuracy is a function of both off-axis position  and source counts (see their Fig. 7 and Table~8). Fig.~\ref{fig:astrom} demonstrates the latter effect, whereby the positions degrade somewhat for fainter sources. For relatively bright X-ray sources ($>100$ net counts) we find 92~per cent of counterparts within 1" and 99~per cent within 2". With $<100$ net counts, we find 78~per cent of the counterparts within 1", but still 96~percent within 2". The RMS error for all sources is 0.83" (0.62" for $>100$ net couthnts and 0.93" for $<100$ counts). 

\subsection{Comparison of associations with G12}
\label{g12_opt}

For the 864 X-ray sources that are in common between our catalogue and that of G12, we have compared the counterpart identifications. There is a substantial difference in the multi wavelength datasets used: G12 used only the DEEP2 optical photometry, while we employ a much wider variety of data including (deblended) IRAC photometry which is known to yield more efficient counterpart identification for X-ray sources (Cardamone et~al. 2008; L09){, where the AGN shine and the number of background sources decrease, making easier the correct and secure association}. Of the 864 common sources, G12 assign a counterpart to 606. By comparison our method yields reliable associations for 830 sources, including 595 of the G12 counterparts. Of the 606 G12 counterparts, our methodology further indicates an unreliable association in 11 cases. 

\begin{figure}
\begin{center}
\hspace{-0.05in}
\epsscale{0.3}
\scalebox{0.4}
{\includegraphics{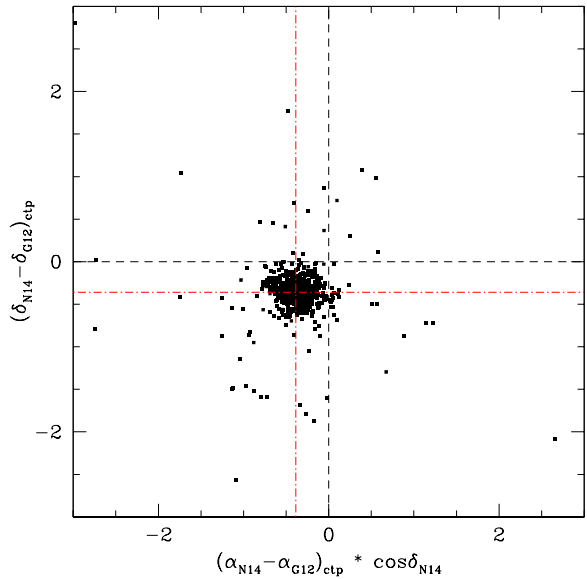}}
\caption{Positional offset between our counterparts and those of G12. A total of 595 sources are assigned both a counterpart in G12, and a secure counterpart in our work. The figures shows a systematic offset between the counterpart positions, presumably associated with the different astrometric frame used for the Rainbow database (our work) and DEEP2 (G12). After correction for this offset the vast majority (572/595 or 97~\%) have the same counterpart within a 0.5" radius. A few show larger differences, indicating that our methodology finds a different counterpart, or has a significantly different position e.g. due to the fact that we perform deblending.
\label{fig:cpt_offset}}
\end{center}
\end{figure}

Matching the counterpart positions, we found a systematic offset between the counterpart positions of about 0.58" (see Figure~\ref{fig:cpt_offset}). This is presumably due to the different astrometric systems adopted by DEEP2 and the Rainbow database.  After correction for this offset we found  572/595 sources with high confidence counterparts in both work to be within 0.5" of each other, and hence are presumably the same counterpart (see Fig.~\ref{fig:cpt_offset}). For the remaining 23 sources we apparently identify a different counterpart to G12. There can be a number of reasons for this, including differences in X-ray positions, deblending or non-deblending of photometry in crowded regions, and the fact that we match to a wider range of catalogues. For example, in several mismatched cases, we identify a single point-like counterpart in the the ACS images that in the ground-based images is merged with 2 or more very nearby objects into an apparent single source, identified as the counterpart by G12. A further check of the consistency between our associations those of G12 can be made by comparing the spectroscopic redshifts of the counterparts, when available. The G12 catalogue contains 180 sources with reliable redshift from DEEP2. For 155 of these, we find the same redshift. For a further two (aegis\_704 and aegis\_762  in this paper) we assign the same counterpart but the redshift is different, as we have adopted a revised redshift value using DEEP3. Of the remaining 23 sources there are 14 sources with spectroscopy in G12 that are not detected as significant X-ray sources in our reduction. For the final 9 we assign a different counterpart to G12, for the reasons discussed above, and hence have a different spectroscopic redshift.



 \section{Redshifts}

The AEGIS field has been the subject of a number of dedicated redshift surveys, most notably the DEEP2 and DEEP3 surveys (Davis et~al. 2003; Cooper et~al. 2011, Cooper et~al. 2012, Newman et~al 2013) with the Keck telescope. The outstanding multi wavelength photometry in the field also permits the determination of accurate photometric redshifts, and the spectroscopic data allow calibration of those redshifts and estimates of their accuracy and reliability. 

\subsection{Spectroscopic redshifts}

A total of 353 sources in our AEGIS-XD X-ray sample have a reliable spectroscopic data, obtained from a variety of sources, listed in priority order  in Table~\ref{tab:spectroscopy}. The largest number (167) of these were derived from the DEEP2 redshift survey (Newman et~al. 2013). DEEP2 was a magnitude-limited redshift survey (to $R_{\rm AB}$=24.1) over several fields covering approximately 2.8 square degrees over 4 fields. Among these fields the AEGIS area features the most extensive multi wavelength coverage (Davis et~al. 2007). Furthermore, the target selection in the AEGIS field was largely based on optical magnitude, unlike the other fields in DEEP2 where color selections were also applied to isolate high redshift galaxies. The redshift success for the AEGIS-XD sources targeted in DEEP2 is very high ($>77$~per cent).  Note that we count only spectra with redshift quality 3 and 4 as defined by Newman et~al. (2013) as secure redshifts, ignoring lower qualities. On the other hand, the sampling rate of DEEP2, combined with selection against presumed stellar objects in the survey means that not all X-ray source counterparts brighter than the magnitude limit were covered. In addition, because the X-ray sources were not known at the time the DEEP2 survey was designed, they could not be targeted explicitly. 

\begin{deluxetable}{ccccc}
\tabletypesize{\scriptsize}
\tablecaption{Spectroscopic redshifts  \label{tab:spectroscopy}}
\tablehead{
\colhead{Survey \tablenotemark{a}} & 
\colhead{$N_{targ}$\tablenotemark{b}} & 
\colhead{$N_{spec}$\tablenotemark{c}} & 
\colhead{$N_{used}$\tablenotemark{d}} & 
\colhead{$N_{star}$\tablenotemark{e}} 
}
\startdata 
DEEP3 &  174  &   89  & 89 &  3 \\
DEEP2 &  223  &  172 & 167  &   0 \\
MMT     &  162  &   93  &   91 & 10  \\
CFRS   &     5  &    5  &    0 & 0 \\
SDSS   &    14 &   14  &  6 &  1 \\
LBG      &  7 & 7 & 6 & 0 \\
\hline
& & &  & \\
 Total & &   464  &  339  &   14 \\
\enddata
\tablenotetext{a}{The redshift origins are listed in priority order, i.e. the redshift is taken preferentially from DEEP3 if available, then DEEP2 and so on down to SDSS, which has the lowest priority. Secure redshifts and stellar identifications are those with redshift quality 3 or 4 in the catalogues. Lower quality flags are assumed to be redshift failures.}
\tablenotetext{b}{Number of AEGIS-XD counterparts targeted for spectroscopy}
\tablenotetext{c}{Secure redshifts and stellar identifications are those with redshift quality 3 or 4 in the catalogues. Lower quality flags are assumed to be redshift failures.}
\tablenotetext{d}{Unique reliable spectroscopic identifications used in this work.}
\tablenotetext{e}{Number of spectroscopically confirmed stars. The difference between $N_{used}$ and $N_{star}$ provides the number of galaxies.}
\end{deluxetable}

The DEEP3 survey (Cooper et~al. 2011), an extension of DEEP2, addressed this issue by specifically targeting the counterparts of  X-ray sources in the field regardless of their optical properties. DEEP3 provides an additional 89 spectroscopic redshifts for the AEGIS-XD survey. The success rate in DEEP3 ($\sim 51$~per cent) was lower than in DEEP2 because targets fainter than the DEEP2 magnitude limit were included. These spectra often failed to yield a secure redshift. A further important spectroscopic dataset was provided using the MMT/Hectospec instrument (Coil et~al. 2009), again explicitly targeting X-ray sources which were not already covered by DEEP2 . This campaign provided a total of 81 secure redshifts for AEGIS-XD sources. Finally, a handful of spectroscopic redshifts of X-ray source counterparts have been obtained by other campaigns, for example the Canada-France Redshift Survey (Lilly et~al. 1995), the Keck Lyman Break Galaxy (LBG)  surveys of Steidel et~al (2003, 2004), and the SDSS (Ahn et~al., 2012). Of these, SDSS provides 6 additional redshifts, and the LBG survey  additional 6. The 5 redshifts from CFRS were all duplicated in DEEP2/3 or the MMT surveys, so we use those in preference. 

\begin{figure}
\begin{center}
\scalebox{0.6}
{\includegraphics{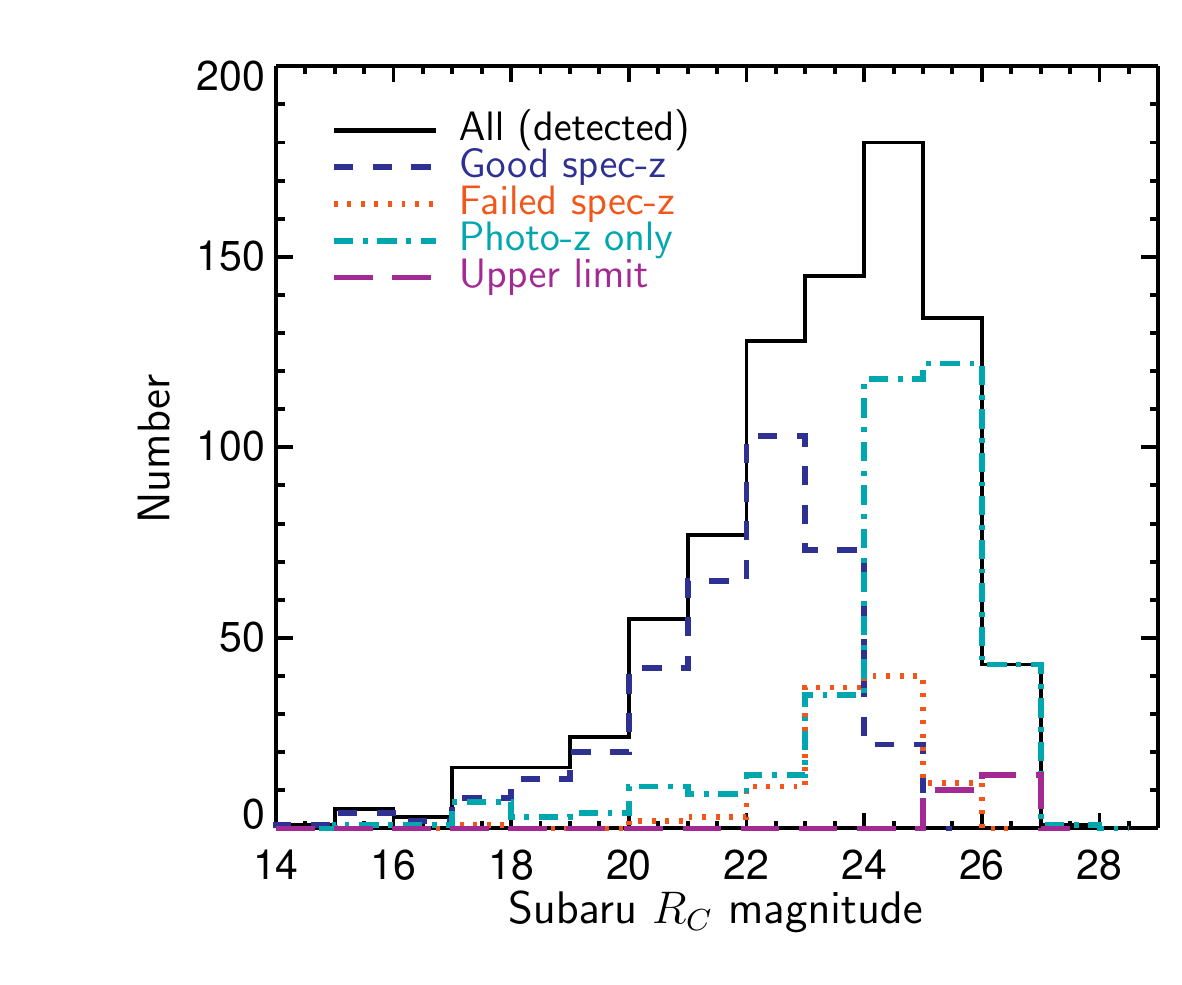}}
\caption{Subaru $R_{C}$ magnitude distribution of the AEGIS-XD counterparts. The distribution peaks at $R_{C}\sim24$. The vast majority of successful spectroscopic redshifts are at brighter magnitudes than this. Redshift failures start to rise sharply at $R_{C}>23$, below which we must largely rely on photometric redshifts. The upper limits curve refers to the upper limit on the magnitude in the case where there is no detected counterpart in the Subaru imaging. \label{fig:zdist}}
\end{center}
\end{figure}

The grand total of 353 spectra implies a total spectroscopic completeness of the sample of $\sim 38$~per cent. This is, however, a very strong function of optical magnitude, as can be seen from Fig.~\ref{fig:zdist}. Only 20 spectroscopic redshifts have been obtained at magnitudes fainter than $R_{C}=24$, despite this being the peak of the magnitude distribution of the X-ray counterparts. This accounts for the relatively low spectroscopic completeness of the whole sample, despite major efforts in terms of spectroscopy in this field. A total of 111 X-ray source counterparts were targeted in the course of the various surveys without a reliable redshift being obtained. 


\begin{figure}
\begin{center}
\hspace{-0.050in}
\vspace{-1in}
\epsscale{0.99}
{\scalebox{0.35}
{\includegraphics{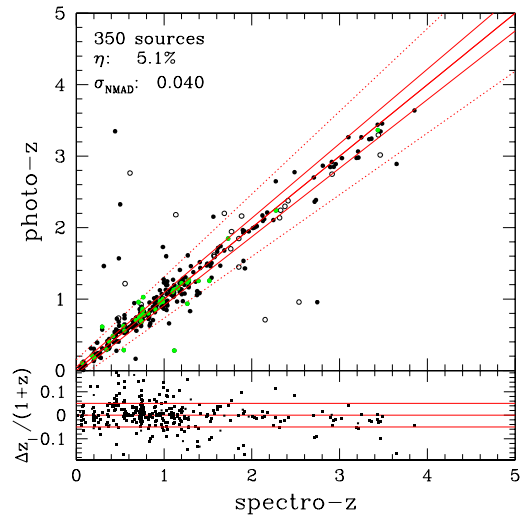}
 }
\caption{Comparison between spectroscopic and photometric redshifts for the extragalactic spectroscopic sample . Filled dots indicate sources with a possible unique redshift solution while 
open circles represent sources for which there is at least a second significant peak in the redshift probability distribution. In green we show the sources within the CANDELS area, where deeper and better resolved NIR data are available, yielding superior photo-z results. The solid lines correspond to z$_{\rm phot}$=z$_{\rm spec}$ and  z$_{\rm phot}=\pm$0.05(1+\rm z$_{\rm spec}$), respectively. The dotted lines limit the locus where z$_{\rm phot}$=$\pm$0.15(1+z$_{\rm spec}$).
 \label{fig:zphot}}
}
\end{center}
\end{figure}


\subsection{Photometric redshifts}

Despite the intensive spectroscopy in this field, a relatively large fraction of the AEGIS-XD sources do not have spectroscopic redshifts. Photometric redshifts were therefore computed for the remaining of the sources with multiwavelength counterparts using SED fitting. 
The photometric redshifts have been computed following the analysis procedure described in Salvato et~al. 2011, which takes into account knowledge of the optical morphology, optical variability and X-ray emission to determine the most appropriate library and priors to be used.  This method has been successfully applied in the COSMOS field (Salvato et~al. 2009), the Lockman Hole (Fotopoulou et~al. 2012) and in the Extended Chandra Deep South (Hsu et~al., 2014). More details regarding the adopted libraries can be found in Hsu et~al 2014.

Morphological classifications for the optical counterparts was taken from the Rainbow survey. There, for the separation of the sources in point-like and extended, the HST-ACS images where primarily used, with the analysis done alternatively on the Subaru $R_{c}$ band image (FWHM$\sim$0.7''), when the former was not available. The motivation for this is that optically-extended sources -- i.e. resolved galaxies - are likely to be better modeled with a galaxy dominated template. Optical point sources and/or variable sources are candidate type 1 QSOs,  better fit with an AGN template.  We classified 401 sources  as optically extended (EXTNV) and the remaining 536 as point-like or unresolved/undetected (QSO variable or QSOV){\footnote{we adopt the same classification as defined in Salvato et~al. (2009) for ease of comparison}.  The EXTNV group was also split on the basis of the soft X-ray flux, with 529 sources fainter than F$_{(0.5-2 \rm keV)} =8 \times 10^{-15}$~erg cm$^{-2}$ s$^{-1}$ and with the remaining 6 brighter than that.

Source variability can be a major  issue in determining the photometric redshifts for type I AGN, as the multi-band data were usually taken at different epochs, meaning that there can be significant flux-offsets worsening the SED fits or causing the wrong template to be selected. In addition, in many cases the photometry in a given band was produced by adding  the results from different observation runs separated in time. This means that a multi-epoch variability analysis was not possible. However, we do find that 59 sources have a clear offset in (typically 0.5 magnitudes and up to $\sim 1$) when comparing photometry from the same or similar filters, suggesting  variability. Visual inspection shows that 38 of these sources are point-like, and while the morphological classification is a strong function of magnitude and image resolution}, this suggests that the variability is likely real and due to the QSO or stellar nature of the source. For the remainder, classified as optically extended, half are located close to nearby stars, and the variability is likely more related to variation in the flux of the stars or the background in their vicinity. For lack of further information, all the 59 {\it apparently-variable} sources have been flagged, as the variability might have a significant effect on the redshift determination. As in Salvato et~al.\, (2011), we used the LePhare code (Ilbert et~al.\, 2006), to compute the photometric redshifts. The first step was to correct the photometry for Galactic extinction using a median E(B-V)=0.04 (see second-last column of  Table \ref{tab:photometry},  as  in Barro et~al.\, 2011a). Then we searched for possible zeropoint offsets that could affect the accuracy of the photometric redshift. To do this we computed the photometric redshift of a sample of normal galaxies (i.e. non X-ray detected) with reliable spectroscopic redshift available. For each source, we kept the redshift fixed and we searched for the best fitting template in the library of  normal galaxies used in Ilbert et~al.\ (2009). Then, for each photometric band, we computed the average difference between the photometry of all the sources and the photometry  of the template. In this way, we can correct for second order problems in the photometric calibration. We adopted these zeropoints (reported in the last column of Table \ref{tab:photometry}) when computing the photometric redshifts for the X-ray selected sources. We note that the zeropoint corrections depend on the templates used, and different libraries could provide slightly different values for the correction. For this reason we do not apply these corrections to the photometric catalogue released with this paper. Furthermore, the same procedure could not be applied directly to the X-ray sources as a) variability could affect the results and b) the relative host/AGN contribution to the SED in a given band is unknown. Ignoring these fact can potentially  introduce a greater uncertainty in the zeropoints when applying a relatively limited number of templates.
 
We compute the photometric redshift for the EXTNV sources using the new hybrid templates of Hsu et~al. (2014). These templates are tuned for sources dominated by galaxies, with special care give in reproducing the emission lines. For the QSOV sample  we used the AGN dominated  hybrids  of Salvato et~al. (2009). Different absolute magnitude priors were considered for  EXTNV ($-24<{\rm M_B}<-8$) and QSOV ($-30<{\rm M_B}<-20$) sources, respectively. In order to search for stars in the sample, a stellar library was also used to fit the all the data. Whenever a source in the QSOV sample is better fit with the stellar template it has been classified as a star. There are 21 sources that satisfy this criterion ($\sim$ 3.1\% of the entire sample), with 12 of these also spectroscopically confirmed. Turning to the extragalactic sources, the overall accuracy for the spectroscopic sample, measured by the normalised mean absolute deviation, is $\sigma_{NMAD}$=0.040, with an outlier fraction of  $\eta$= 5.1\% (Fig.~\ref{fig:zphot}). The accuracy is slightly higher ($\sigma_{NMAD}$=0.030, with an outlier fraction of  $\eta$= 3.8\%, where deeper and better resolved NIR photometry from CANDELS is available). A close look to the outliers revealed that most of those for which the photometric redshift solution  was not unique (empty circles in the figure), had the correct solution in the second higher peak. The use of  redshift probability distribution function (made publicly available with this work) is recommended. The remaining outliers can be explained by blending with nearby sources for which the low resolution of the ground-based imaged misplace the sources in the "EXTNV" group, rather than in the "QSOV", thus adopting the wrong templates/priors.  The effect of the resolution of the images used for the classification has been addresses in Hsu et~al. (2014). These authors demonstrate that  the fraction of outliers increased significantly when using the morphological classification from the ground-based images rather than from the space-based ones. We  break down the results in redshift, magnitude and type in Table \ref{tab:zphot_accuracy}.  The outlier fraction and uncertainties in the QSOV sample are larger, partly due to degeneracies associated with the typical power-law SED of this subsample. This is a general problem for photometric redshift for AGN, which can be mitigated when narrower filters  sensitive to strong emission lines are used in the photo-z determination (e.g. Salvato et~al. 2009; Cardamone et~al. 2010; Hsu et~al. 2014). 

The full probability distribution function for the photometric redshifts is made available for all the sources in our catalogue (see Appendix \ref{sec:dr}). We caution that the errors associated with the photometry can easily be underestimated, with the result that the 1-3 $\sigma$ errors associated with the photometric redshift can also be underestimated. This is true for photometric redshift in general (see  Dahlen et~al.\, 2013) for a recent test with a sample of normal galaxies in the CANDELS fields). We verified that the situation is similar for our sample where  z$_{\rm phot}-1\sigma < z_{\rm spec} < z_{\rm phot}+1\sigma$ only for 57\%  of the sources, while for 79\% of the sample the spectroscopic sample is within the 3$\sigma$ error of the photometric redshift. Thus, while using the 3$\sigma$ errors associated with photometric redshifts provide a reasonably accurate estimate of the uncertainty for most of the sources, we recommend again to use the entire redshift probability distribution function  for a more complete analysis.



\begin{table*}
\caption{Accuracy of photometric redshifts. (1) Number of sources (2) Outlier fraction (3) Normalised mean absolute deviation. QSOV refers to optically point-like and/or variable sources. EXTNV refers to optically extended and non-variable sources.}
 \begin{center}
 
 \begin{tabular}{c|ccc|ccc}

\hline
                                  Sub Sample  &  \multicolumn{3}{c}{QSOV} &    \multicolumn{3}{|c}{EXTNV}\\
& N. of sources & $\eta$ (\%) &$\sigma_{NMAD}$  & N. of sources &$\eta$ (\%) &$\sigma_{NMAD}$  \\  
& (1) & (2) & (3) & (1) & (2) & (3) \\ 
\hline
\hline
z$<$1.2     &    53     & 11.3 & 0.067 & 201 & 3.5 & 0.034 \\
z$>$1.2              &72& 5.6 & 0.059 & 24 & 4.2 & 0.027\\
\hline
R$<$22 mag             & 41 & 7.3 & 0.067   & 111 & 2.7 & 0.040 \\
R$>$22 mag          &84&8.3&0.061& 114& 4.4 & 0.028\\
\hline
 Combined & 125&8.0&0.063&225&3.6&0.033\\
\end{tabular}
\end{center}
 \label{tab:zphot_accuracy}
\end{table*}


\begin{figure}
\epsscale{0.7}
{\scalebox{0.4}
{\includegraphics{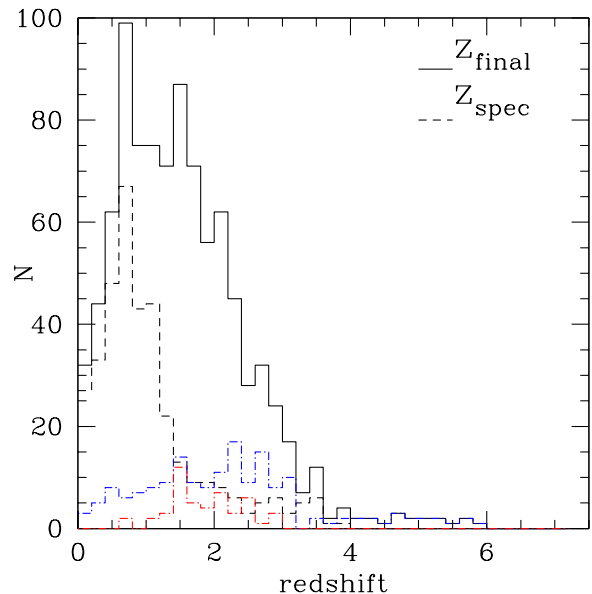}
}
\caption{Redshift distribution for the X-ray sources. The black dashed line indicates the distribution for the sources with secure spectroscopic redshift  and in black solid line the same distribution is  added to the rest of the sources for which only a photo-z is available.
In red we indicate the sources for which there are at least two peaks in the redshift solution while in blue we show the distribution of the sources for which there is a single peak, but with a low value of P(z). \label{fig:z_histogram}}
}
\end{figure}


 \subsection{Redshift distribution}
 \label{sec:highz}

The redshift distribution of our X-ray sources (after excluding the  stars) is shown in  Figure \ref{fig:z_histogram}, distinguishing between spectroscopic and photometric redshifts. There is a peak in the spectroscopic redshift distribution around $z\sim$0.7, which  is also seen in the photometric redshift distribution, presumably due to a large-scale structure at this redshift in the field. The redshift distribution shows clearly that the spectroscopic completeness is a very strong function of redshift, with the sample being highly spectroscopically complete at $z<1$ (210/303), and highly incomplete above this redshift  (140/613). This shows the great importance of computing accurate photo-z when considering AGN evolution and/or host galaxy properties of high redshift. 

Figure \ref{fig:z_histogram}  also shows  in blue the sources that are characterized by at least one further peak in the redshift solution (200 sources, 23\% of the sample) and in red sources for which a single peak in the redshift solution is found but for which the P(z) is lower than 50\% (34 sources, 6\%), with a distribution over a broad redshift range.
The majority of the high redshift sources (i.e. z$>$ 3.5) are among these poorly defined sources.  We find 14 X-ray sources at $z=3.5-4.5$ in the AEGIS field (2 spectroscopically confirmed), and 15 at redshift z$>4.5$, although all of the latter show a secondary peak at a lower redshift. In this regime the photometric redshifts are strongly dependent on the upper limits adopted, making the results unstable. The high redshift nature of these sources thus needs to be assessed carefully, ideally via spectroscopy or possibly with deeper photometry (e.g. Venemans et~al. 2007).  For the time being the photometric redshifts should be considered with the associated errors and preferably with the full P(z).





\begin{figure*}
\hspace{-0.05in}
\epsscale{0.8}
{\scalebox{0.4}
{\includegraphics{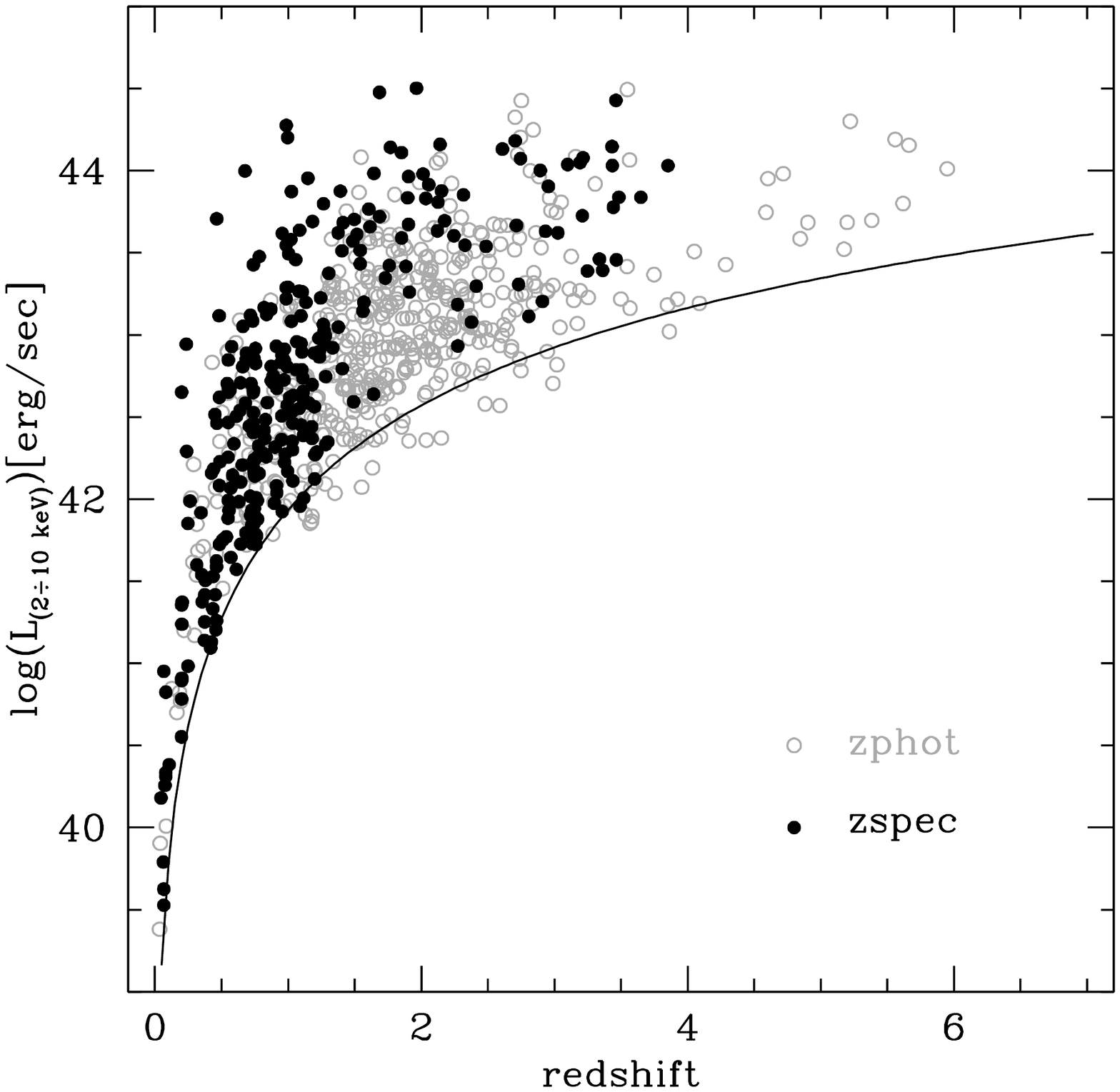}
\includegraphics{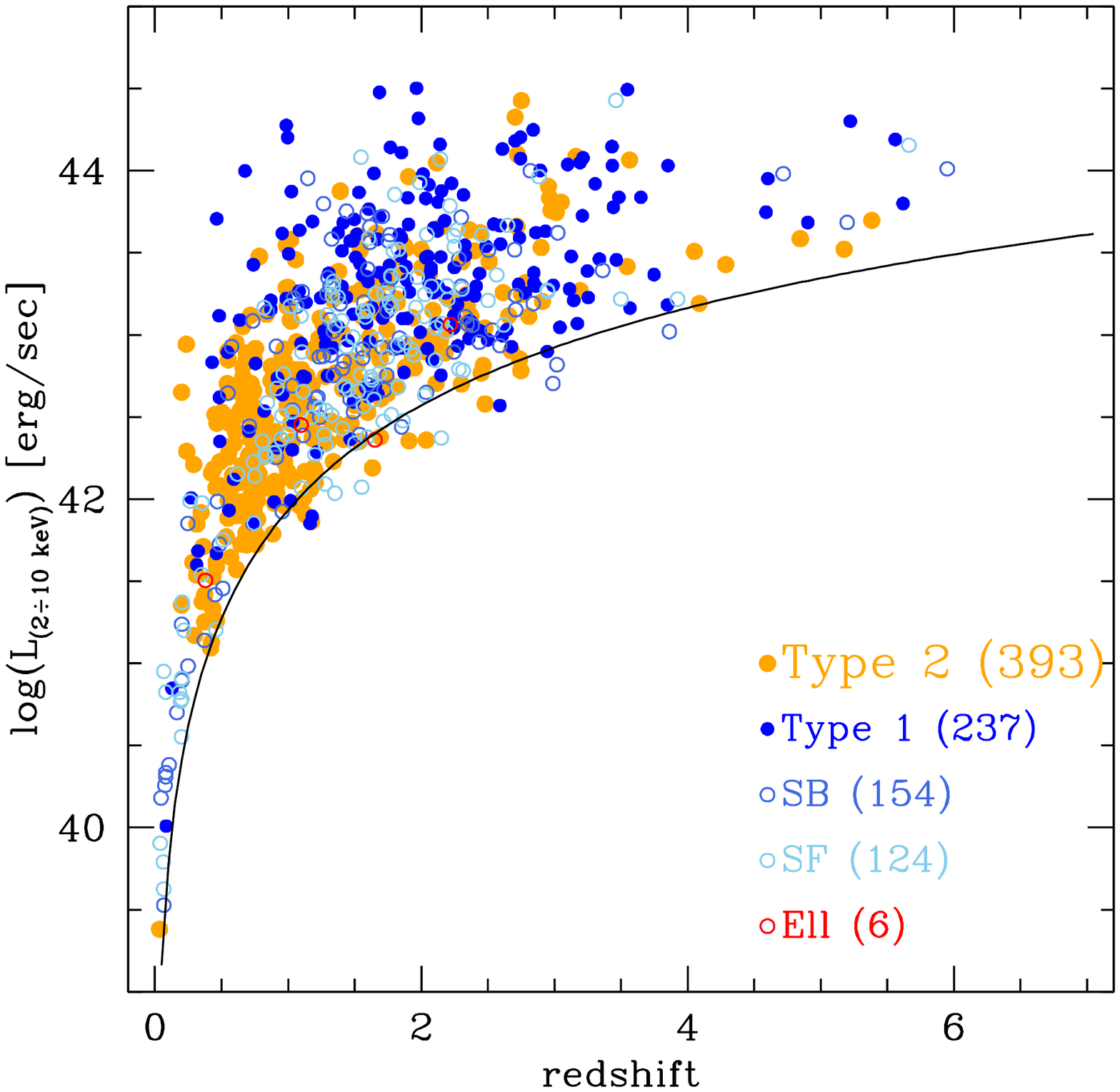}
}
\caption{Left panel: X-ray luminosity in the  rest frame 2-10 keV as a function of spectroscopic (black filled circles) or photometric redshift (gray open circles). Right panel: same plot but this time the sources are color coded on the basis of the best-fit template to the photometric data. Here filled symbols indicate sources best-fit by an AGN template (type 1 or 2), or a hybrid with some contribution (from 10 to 100\%) from an AGN. Galaxy templates are shown as open symbols and include Ellipticals (Ell), various spiral/irregular templates (SF) and starbursts (SB). The number of sources of each SED type is also reported in the figure. Note that  the sources fit by a AGN/hybrid are on average more luminous in the X-ray, as might be expected. Even if the optical/IR SED is better fit with a galaxy template, however, the X-rays will in the vast majority of cases be dominated by an AGN.  In both figures the black solid curve corresponds to a flux limit of $f_X(\rm 2-10\,keV) = 2.5 x 10^{-16} erg\,  s^{-1}  \, cm^{-2}$, where the 2 -10\,keV sensitivity curve drops to 1\% of the maximum  value.  For the calculation of 2-10 keV luminosity we used the 0.5-10 keV flux, k-corrected using a power-law X-ray spectrum with spectral index $\Gamma=1.4$. Because the 0.5-10 keV band is more sensitive than the 2-10 keV band, a few sources appear below the line corresponding to the flux limit. \label{fig:Lx_z_type}}
}
\end{figure*}

Fig.~\ref{fig:Lx_z_type} shows our sample in the $L_{\rm X}-z$ plane. The left panel distinguishes between sources with spectroscopic redshifts and those with photometric redshifts. This clearly shows that that vast majority of sources at $z>1.5$ do not have spectroscopic redshift, making the photo-z essential. Obtaining accurate phot-z, as in our work, is thus crucial for the investigation of AGN properties at high redshift. The AEGIS survey alone provides reasonably good sampling of the population in the luminosity range $10^{42-44}$ up to around $z~3$, after which is becomes incomplete at the faint end. This can be improved by combining it with deeper surveys such as the CDF-N and, particularly, the CDF-S. Similarly, the luminosity range above $10^{44}$~erg s$^{-1}$, above $L_{*}$, the knee in the XLF, is relatively sparsely sampled by our data and larger area surveys are required e.g. to determine the bright end of the XLF with good accuracy. The right panel of Fig.~\ref{fig:Lx_z_type} shows the objects colour-coded by the best fit template fitted to the multi-wavelength photometry during the photo-z determination. There is a considerable mix of template types, depending on whether the light is dominated by the AGN, the galaxy, or a mixture at longer wavelengths.  Previous work using the same photometric redshift methods as use here has shown good agreement between the SED type and the spectroscopic classification, where available (Salvato et al. 2009; Lusso et al.~ 2010, Lusso et al.~ 2012, Civano et al.~2012) 




\section {Summary}
\label{sec:conc}

A catalogue of X-ray sources detected in the deep (800ks)  {\it Chandra} imaging of the AEGIS field has been presented. This is currently the third deepest X-ray survey in existence, after the Chandra Deep Fields North and South. A total of 937 X-ray sources have been detected down to a Poisson false probability of $4 \times 10^{-6}$, calculated based on the counts detected in a PSF-sized detection cell and a local background, following the methodology of L09. The source detection algorithm enables an accurate determination of the sensitivity of the observations over the field using the technique of Georgakakis et~al. (2008b), and hence the catalogue can be used effectively when statistical investigations requiring corrections for completeness are needed (e.g. luminosity functions). We have identified multi-wavelength counterparts to our sources using a wide variety of complementary data in this field, ranging from the UV to the mid-infrared. Using a likelihood-based association method, we find possible counterparts for 929/937 or $\sim 99$~per cent of the X-ray sources, the vast majority from the deep Spitzer/IRAC imaging in the field. We note, however, that the statistically reliability of likelihood-based associations is not 100~\%, so the notional completeness of the counterpart identifications is closer to $94$~per cent in a statistical sense. 353 ($\sim 38$~per cent) of the X-ray source counterparts have a reliable spectroscopic redshift mostly from Keck spectroscopy in the DEEP2 and DEEP3 surveys, supplemented by a significant number from MMT/Hectospec spectroscopy. For all X-ray source associations, we have performed multi-wavelength photometry in up to 35 bands using the methodology pioneered in the {\it Rainbow} database (Barro et~al. 2011). This provides SEDs for the sources and furthermore enables accurate photometric redshift to be determined, using the methodology of Salvato et~al. (2011), which is tuned particularly for X-ray sources detected in deep surveys, which mostly comprise AGN. Despite greater difficulties and uncertainties associated with determining photo-z for such sources, the reliability and accuracy of the photometric redshifts is excellent, with an outlier fraction of just $\eta = 5\%$ and $\sigma=0.05$. Even better results $\eta =4\%$ and $\sigma=0.03$ is reached in the CANDELS area where deeper and superior NIR data are available. The AEGIS-XD dataset lies in a unique area of parameter space in terms of deep X-ray surveys and the excellence of the redshift determinations and the supporting multiwavelength data make it a powerful tool to investigate the AGN population. The dataset described here as already been used to investigate the colours of AGN hosts (Georgakakis et~al. 2014a), AGN clustering (Georgakakis et~al. 2014b), Compton thick AGN (Brightman et~al. 2014) and the evolution of AGN obscuration (Buchner et~al. 2014), typically in combination with other deeper and/or wider datasets. Further work investigating these and related phenomena should greatly enhance our knowledge of black hole growth over cosmic time, and its relationship to galaxy evolution. 

All of our catalogues, including the detailed X-ray information, multi wavelength identifications, aperture-matched photometry and redshift information (including the SED fits and photometric redshift $p(z)$ are released publicly, as described in the Appendix to this paper. 



\acknowledgments
We thank those who have built and operate the \chandra\ X-ray observatory so successfully. We acknowledge financial support from Chandra grant G08-9129A, NSF grant AST-0808133, US National Science Foundation via grant AST-0806732. PGP-P acknowledges support from the Spanish Programa Nacional de Astronom\'{\i}a y Astrof\'{\i}sica under grant AYA2012-37727. This work  has made use of the {\it Rainbow} Cosmological Surveys Database, which is  operated by the Universidad Complutense de Madrid (UCM) partnered  with the University of California Observatories at Santa Cruz (UCO/Lick,UCSC).  Facilities: \facility{CXO(ACIS)}, \facility{CFHT}, \facility{Spitzer(IRAC)}.

\clearpage

\appendix
\section{Catalogues and Data Release}
\label{sec:dr}

The catalogues described in this paper are available via the journal and will be updated whenever new data will be available at the public websites at MPE \url{www.mpe.mpg.de/XraySurveys}  and  via {\it Rainbow} (\url{http://rainbowx.fis.ucm.es/Rainbow\_Database/Home.html}).  Specifically, we release the following products:

\begin{itemize}
\item The X-ray source catalogues.
\item The optical/NIR/MIR association catalog listing the X-ray sources with corresponding multi-band photometry.
\item The redshifts, including spectroscopic redshifts, photometric redshifts and photometric redshift probability distributions.
\end{itemize}

In this appendix, we show extracts from each of the catalogues, with the information provided in the electronic edition of the journal. The full information is provided on the web site in FITS format. 

\subsection{X-ray catalogs} 
\label{sec:xcat}

A subset of the basic X-ray properties of the AEGIS-XD sources are given in Table~\ref{tab:cat1} and in Table~\ref{tab:cat2}. In the first Table we provide co-ordinates of the sources and detection properties in the various X-ray bands. In the second we provide X-ray properties such as fluxes and hardness ratios.   Each table is fully described in the corresponding notes.

\subsection{Multi-wavelength catalog}
\label{catalog:optical}
The multi wavelength properties of the counterparts are  provided in Table~\ref{tab:ctp} and Table~\ref{tab:photometry_example}. In the first Table we provide the coordinates of the counterparts in the optical and near/mid-infrared catalogs, together with their ID and offset from the X-ray coordinates. In the second table, for each of the counterparts we provide the magnitude in the AB system for all the bands listed in the Table \ref{tab:photometry}.
The photometry is corrected for Galactic extinction but is not corrected for zero point offset as this  value is dependent on the SED used for its computation.
In Table~\ref{tab:photometry_example} an excerpt  from the photometric catalog is shown.

\subsection{Redshift catalog}
\label{catalog:redshift}

For each X-ray source we list the spectroscopic redshift where available, the origin of the redshift, and the redshift quality flag.
In addition, we provide the photometric redshift values and their associated 1 and 3 sigma, including possible second solution when appropriate.
LePhare also provides the full redshift probability distribution, P(z), which is available on request. In the current catalog we provide the peak value of P(z) for the first and second solutions, As is visible in the excerpt of Table~\ref{tab:photoz} the peak P(z) is related to the number of photometric points available for the fit, with lower value of the peak associated with sources with fewer photometric points available. This should be borne in mind when assessing the reliability of any given photometric redshift.  
The Table also shows the templates that were selected for the best fit, which are publicly available at
\url{http://www.mpe.mpg.de/XraySurveys}.


\begin{deluxetable}{ll cccc cc cc cc cc c}
\tabletypesize{\tiny}
\rotate
\tablecaption{\chandra\ \ax\ source catalog: basic source properties\label{tab:cat1}}
\tablewidth{0pt}
\tablehead{
\colhead{} &      
\colhead{} &      
\colhead{RA} &            
\colhead{Dec} &           
&         
&         
\multicolumn{2}{c}{FB cts} & 
\multicolumn{2}{c}{SB cts} & 
\multicolumn{2}{c}{HB cts} & 
\multicolumn{2}{c}{UB cts} & 
\colhead{Detection} \\ 

\colhead{ID} &            
\colhead{IAU Name} &       
\colhead{(J2000)} &       
\colhead{(J2000)} &       
\colhead{Pos. err} &      
\colhead{OAA} &   
\colhead{N} &     
\colhead{B} &     
\colhead{N} &     
\colhead{B} &     
\colhead{N} &     
\colhead{B} &     
\colhead{N} &     
\colhead{B} &     
\colhead{bands} \\ 
\colhead{(1)} &
\colhead{(2)} &
\colhead{(3)} &
\colhead{(4)} &
\colhead{(5)} &
\colhead{(6)} &
\colhead{(7)} &
\colhead{(8)} &
\colhead{(9)} &
\colhead{(10)} &
\colhead{(11)} &
\colhead{(12)} &
\colhead{(13)} &
\colhead{(14)} &
\colhead{(15)}\\
}
\startdata
aegis\_001 & AEGISXD J141805.22+522510.6 & 214.521757 & 52.419615 & 0.87 & 12.11 & 125 & 72.99 & 57 & 21.72 & 68 & 53.27 & 43 & 33.92 & FSH \\
aegis\_002 & AEGISXD J141807.03+522523.0 & 214.529298 & 52.423079 & 0.57 & 11.90 & 245 & 76.40 & 105 & 21.80 & 135 & 54.80 & 61 & 34.16 & FSHU \\
aegis\_003 & AEGISXD J141816.27+522524.7 & 214.567822 & 52.423529 & 0.87 & 11.97 & 128 & 43.68 & 58 & 12.46 & 71 & 32.46 & 36 & 20.80 & FSH \\
aegis\_004 & AEGISXD J141826.59+522602.0 & 214.610827 & 52.433892 & 0.87 & 11.68 & 142 & 86.52 & 41 & 24.20 & 105 & 64.69 & 55 & 40.69 & FH \\
aegis\_005 & AEGISXD J141813.90+522625.0 & 214.557935 & 52.440278 & 0.87 & 10.93 & 175 & 123.78 & 54 & 32.87 & 124 & 94.25 & 72 & 60.56 & FS \\
aegis\_006 & AEGISXD J141757.02+522631.3 & 214.487603 & 52.442038 & 0.57 & 10.85 & 183 & 52.06 & 66 & 14.05 & 120 & 40.54 & 49 & 26.50 & FSHU \\
aegis\_007 & AEGISXD J141821.24+522655.7 & 214.588520 & 52.448820 & 0.87 & 10.61 & 232 & 157.02 & 77 & 42.63 & 160 & 119.96 & 94 & 77.03 & FS \\
aegis\_008 & AEGISXD J141822.87+522709.5 & 214.595307 & 52.452642 & 0.87 & 10.44 & 240 & 154.97 & 128 & 42.36 & 106 & 117.96 & 72 & 76.39 & FS \\
aegis\_009 & AEGISXD J141829.77+522709.6 & 214.624081 & 52.452675 & 0.87 & 10.75 & 163 & 100.76 & 44 & 27.15 & 127 & 77.76 & 71 & 49.70 & FH \\
aegis\_010 & AEGISXD J141804.84+522740.2 & 214.520174 & 52.461183 & 0.57 & 9.61 & 213 & 105.52 & 93 & 27.81 & 118 & 81.69 & 66 & 54.83 & FS \\

\enddata
\tablenotetext{1}{Unique source name}
\tablenotetext{2}{IAU source name}
\tablenotetext{3}{X-ray position RA in degrees}
\tablenotetext{4}{X-ray position Dec in degrees}
\tablenotetext{5}{X-ray positional error in arcsec}
\tablenotetext{6}{Off axis angle in degrees.}
\tablenotetext{7}{0.5-7\,keV (full band) total counts extracted at the source position within the 90\% EEF radius.}
\tablenotetext{8}{Background counts in the 0.5-7\,keV band scaled to the area that corresponds to the 90\% EEF radius.}
\tablenotetext{9}{0.5-2\,keV (soft band) total counts extracted at the source position within the 90\% EEF radius.}
\tablenotetext{10}{Background counts in the 0.5-2\,keV band scaled to the area that corresponds to the 90\% EEF radius.}
\tablenotetext{11}{2-7\,keV (hard band) total counts extracted at the source position within the 90\% EEF radius.}
\tablenotetext{12}{Background counts in the 2-7\,keV band scaled to the area that corresponds to the 90\% EEF radius.}
\tablenotetext{13}{5-7\,keV (ulta-hard band) total counts extracted at the source position within the 90\% EEF radius.}
\tablenotetext{14}{Background counts in the 5-7\,keV band scaled to the area that corresponds to the 90\% EEF radius.}
\tablenotetext{15}{Spectral bands that the source is detected with Poisson background fluctuation probability $<4\times10^{-6}$. The letters correspond to full-band (F), soft-band (S), hard-band (H) and ultra-hard band (U)}

\tablecomments{Table \ref{tab:cat1} is published in its entirety in the electronic edition of the journal.}
\end{deluxetable}
 

\begin{deluxetable}{l cccc c c}
\tabletypesize{\scriptsize}
\tablecaption{\chandra\ \ax\ source catalog: source fluxes and HRs\label{tab:cat2}}
\tablewidth{0pt}
\tablehead{
\colhead{} &      
\multicolumn{4}{c}{Bayesian flux} & 
\colhead{Bayesian}  &      
\colhead{Phot.}\\      
\cline{2-5}
\colhead{ID} &            
\colhead{$f_{0.5-10}$}&   
\colhead{$f_{0.5-2}$} &   
\colhead{$f_{2-10}$}  &   
\colhead{$f_{5-10}$}  &   
\colhead{HR}          &   
\colhead{flag}        \\  
\colhead{(1)} &
\colhead{(2)} &
\colhead{(3)} &
\colhead{(4)} &
\colhead{(5)} & 
\colhead{(6)} &
\colhead{(7)} \\
}
\startdata
aegis\_001 & $29.67^{+6.97}_{-6.37}$ & $8.76^{+2.14}_{-1.87}$ & $15.34^{+9.68}_{-8.56}$ & $<25.33$ & $-0.41^{+0.25}_{-0.24}$ & 0 \\
aegis\_002 & $87.77^{+8.68}_{-8.14}$ & $18.96^{+2.57}_{-2.33}$ & $75.91^{+11.98}_{-10.98}$ & $40.82^{+13.47}_{-11.84}$ & $-0.04^{+0.10}_{-0.10}$ & 0 \\
aegis\_003 & $55.39^{+8.11}_{-7.42}$ & $13.10^{+2.49}_{-2.18}$ & $45.96^{+11.29}_{-10.02}$ & $29.08^{+13.51}_{-11.42}$ & $-0.10^{+0.14}_{-0.13}$ & 0 \\
aegis\_004 & $28.98^{+6.76}_{-6.22}$ & $3.87^{+1.72}_{-1.47}$ & $37.81^{+10.58}_{-9.59}$ & $<24.85$ & $0.37^{+0.20}_{-0.18}$ & 0 \\
aegis\_005 & $17.62^{+4.90}_{-4.55}$ & $3.19^{+1.27}_{-1.11}$ & $18.44^{+7.54}_{-6.89}$ & $<19.55$ & $0.11^{+0.30}_{-0.20}$ & 0 \\
aegis\_006 & $60.44^{+6.72}_{-6.24}$ & $10.44^{+1.84}_{-1.63}$ & $67.13^{+10.13}_{-9.24}$ & $30.36^{+10.87}_{-9.41}$ & $0.20^{+0.10}_{-0.10}$ & 0 \\
aegis\_007 & $22.31^{+4.84}_{-4.53}$ & $4.53^{+1.30}_{-1.15}$ & $21.30^{+7.28}_{-6.72}$ & $<18.92$ & $0.02^{+0.22}_{-0.18}$ & 0 \\
aegis\_008 & $24.74^{+4.81}_{-4.50}$ & $11.03^{+1.59}_{-1.46}$ & $<14.08$ & $<18.24$ & $-0.88^{+0.03}_{-0.12}$ & 0 \\
aegis\_009 & $25.82^{+5.72}_{-5.29}$ & $3.10^{+1.42}_{-1.22}$ & $36.46^{+9.11}_{-8.33}$ & $25.49^{+11.34}_{-10.06}$ & $0.45^{+0.19}_{-0.16}$ & 0 \\
aegis\_010 & $31.84^{+4.63}_{-4.32}$ & $8.46^{+1.39}_{-1.25}$ & $19.52^{+6.40}_{-5.83}$ & $<16.54$ & $-0.31^{+0.16}_{-0.14}$ & 0 \\

\enddata
\tablenotetext{1}{Unique source name}
\tablenotetext{2}{Flux in the 0.5-10\,keV band in units of $10^{-15} \, erg \, s^{-1}\, cm^{-2}$ estimated using the Bayesian methodology of L09}
\tablenotetext{3}{Flux in the 0.5-2\,keV band in units of $10^{-15} \, erg \, s^{-1}\, cm^{-2}$ estimated using the Bayesian methodology of L09}
\tablenotetext{4}{Flux in the 2-10\,keV band in units of $10^{-15} \, erg \, s^{-1}\, cm^{-2}$ estimated using the Bayesian methodology of L09}
\tablenotetext{5}{Flux in the 5-10\,keV band in units of $10^{-15} \, erg \, s^{-1}\, cm^{-2}$ estimated using the Bayesian methodology of L09}
 
\tablenotetext{6}{Hardness ratio determined by BEHR (Park et~al. 2006) using the counts in the 0.5-2 and 2-7\,keV spectral bands}
 
\tablenotetext{7}{Quality of the X-ray photometry. A flag of ``1'' indicates the presence of a nearby
source that may be contaminating the photometry. A flag of ``2'' indicates that another source was detected with the 90\% EEF
and that the photometry is likely heavily contaminated and the source position uncertain. All other sources have a
flag of ``0''.}
\tablecomments{Table \ref{tab:cat2} is published in its entirety in the electronic edition of the journal.}
\end{deluxetable}


\begin{deluxetable}{ccccccccc}
\tabletypesize{\scriptsize}
\tablecaption{Optical and IR counterparts to the AEGIS-X sources. \label{tab:ctp}}
\tablewidth{0pt}
\tablehead{
\colhead{XID \tablenotemark{a}} & 
\colhead{OBJ\_NO\tablenotemark{b}} & 
\colhead{AEGIS ID (Rainbow)\tablenotemark{c}} & 
\colhead{X-ray R.A.\tablenotemark{d}} & 
\colhead{X-ray Dec.\tablenotemark{d}} & 
\colhead{Ctrp. R.A.\tablenotemark{e}} &
\colhead{Ctrp. Dec. \tablenotemark{e}} &
\colhead{PRIM\_MATCH \tablenotemark{f}} & 
\colhead{ROBUST\_Ctrp. \tablenotemark{g}} \\
\colhead{} &
\colhead{} &
\colhead{} &
\colhead{(J2000)} & 
\colhead{(J2000)} & 
\colhead{(J2000)} & 
\colhead{(J2000)} & 
\colhead{} &  
\colhead{}   
}
\startdata

 aegis\_293  & 533   &    aegis\_293\_1 &  214.448561  &  52.695391 &  214.4488572  &  52.6953656 &   SubaruR &     1 \\
  aegis\_901  & 1620  &   aegis\_901   &  215.225446  &  53.118118  & 215.2257595   & 53.1182402   & RAINBOW  &    1 \\
 aegis\_819   &1488   &  aegis\_819   &  214.769967  & 53.047322  & 214.7703197   & 53.047412    & CFHTLS      & 1 \\
 aegis\_291  & 531  &    aegis\_291    &  214.409987  & 52.693303  & 214.4098638  &  52.6932229   & RAINBOW   &   1 \\
 aegis\_418  & 762   &   aegis\_418    &  214.585343  & 52.788107  & 214.5855658   & 52.7883566   & RAINBOW    &  1 \\
 aegis\_355  & 647    &  aegis\_355    &  214.959983   & 52.743431  & 214.9594025   & 52.7435032   & RAINBOW    &  1 \\
 aegis\_486   & 894    &  aegis\_486    &  214.723504  & 52.839885  & 214.7237642  &  52.8398494  &  SubaruR   &   1 \\
 aegis\_529   & 962     & aegis\_529    & 214.792841  & 52.86416    & 214.7927908   & 52.864219    & RAINBOW   &   1 \\
 aegis\_296   & 538    &  aegis\_296   &  214.529397  & 52.697265  & 214.5293663   & 52.6971555   & RAINBOW   &   1 \\
 aegis\_541   & 980    &  aegis\_541   &  214.671679  & 52.871095   & 214.6720393   & 52.8710399   & RAINBOW   &   1 
 
\enddata
\tablecomments{Table \ref{tab:ctp} is published in its entirety in the electronic edition of the journal.}
\tablenotetext{a}{X-ray Identification}
\tablenotetext{b}{Object number in Rainbow}
\tablenotetext{c}{Object Identification in Rainbow}
\tablenotetext{d}{X-ray position}
\tablenotetext{e}{Counterpart position}
\tablenotetext{f}{Band first used to identify multi wavelength counterpart. "none" indicate no counterpart identified}
\tablenotetext{g}{Flag indicating whether the association with a counterpart is secure (1) or not (0).}
\end{deluxetable}

\begin{deluxetable}{lccccccccc}

\tabletypesize{\scriptsize}
\tablecaption{Extract from the AEGIS-XD  Photometric  catalog \label{tab:photometry_example}}
\tablewidth{0pt}
\tablehead{

\colhead{XID} &
\colhead{AEGIS ID ({\it Rainbow})} & 
\colhead{RA$_{opt}$} &
\colhead{DEC$_{opt}$} &
\colhead{R} &
\colhead{err$_{R}$} &
\colhead{...} &
\colhead{...} &
\colhead{FUV} &
\colhead{err$_{FUV}$} \\
\colhead{(1)} &
\colhead{(2)} &
\colhead{(3)} &
\colhead{(4)} &
\colhead{(5)} &
\colhead{(6)} &
\colhead{(...)} &
\colhead{(...)} &
\colhead{(73)} &
\colhead{(74)} 
}
\startdata
549 & aegis\_293\_1 &  214.4488572 &  52.6953656 & 14.86 & 0.02 &...&...& 22.027 & 0.03 \\
551 & aegis\_294  & 214.4789369 &  52.6955591 &21.94 &0.02 &...&...&-99.0 &-99.0 \\
598 & aegis\_321           & 214.9992079 & 52.7125128 &20.69 &0.02 &...&...&-99.0 &-99.0 \\
749 & aegis\_399\_1     & 214.9281189 & 52.7771454 &21.9 &0.02 &...&...&-99.0& -99.0 \\
1195 & aegis\_626       &214.7869854  & 52.9435878 &23.2 &0.03&...&...& -99.0& -99.0 \\
1276 & aegis\_669      & 215.198185    & 52.969059   & 18.62 & 0.02 &...&...&22.627 &0.03 \\
6        & aegis\_004\_1       & 214.6103281  & 52.4330603    & 24.65 &0.05 &...&...&-99.0 &-99.0 \\
11 & aegis\_005\_4     &214.55811309  & 52.44051303 & 25.27 &0.08 &...&...&-99.0 &-99.0 \\
16 & aegis\_008         & 214.5951844   & 52.4524551    & 17.69 &0.02 &...&...&24.147& 0.17 \\
17 & aegis\_009\_1   &214.6241732    & 52.4526771    & 22.6 &0.02 &...&...&-99.0 & -99.0
\enddata
\tablecomments{Excerpt from the photometric catalog. Column 1: X-ray ID; Column 2: Optical identifier number from the {\it Rainbow} catalog; Columns 3 and 4;  Right Ascension and Declination in degrees of the counterpart; Column 5 and following odd columns : AB magnitude in the filters listed in Table \ref{tab:photometry}.;  Column 6 and following even columns: associated photometric errors.}
\end{deluxetable}

 
\begin{rotate}{90}
\begin{table}[htdp]
\label{tab:photoz}

\caption{Extract from the AEGIS-XD redshift catalog}

\begin{center}
\begin{tabular}{cccccccccccccccc}

\small XID &  z$_{spec}$    & z$_{conf}$& z$_{spec}$ Ref. &N$_{bands}$ &z$_{p1}$   & z$_{p}$L 1$\sigma$ & z$_{p1}$U  1$\sigma$& z$_{p1}$L 3$\sigma$ & z$_{p1}$U 3$\sigma$   & P(z$_{p1}$) & Mod$_1$ & z$_{p2}$ &P(z$_{p2}$) & Mod$_2$  \\
            (1)  &    (2)         &   (3)                &     (4)                    &    (5)                      &  (6)     &     (7)       &    (8)   &      (9) &(10)&(11)&(12)&(13)&(14)   &(15)  \\
 \hline
 \hline

549   & 0.066236     &4& 2& 25&0.0419 &  0.02 &  0.060      &0.020 &0.070 &99.999 &107& -99.0 &0.0 &-999\\
551   &0.463579     &4 &1 &26&0.557  &  0.490 &  0.630      &0.460 &0.660 &97.877 &124 &-99.0 &0.0 &-999\\
598   &-1.0                &0& 0& 15&0.571 &  0.460 &  0.690      &0.250 &0.720 &74.767 &130 &-99.0 &0.0 &-999\\
749   & 0.784325    & 4&1 &21&0.745 &  0.710 &  0.790       &0.690 &0.810& 98.281& 128 &-99.0& 0.0 &-999\\
1195 &1.392565     & 4 &1&17&1.096 & 1.030 & 1.180        &0.970 &1.270 &89.297 &129 &-99.0 &0.0 &-999\\
1276 & 0.20079   & 4 &1& 20&0.260& 0.200     &0.310          &0.170 &0.340 &95.092 &124 &-99.0 &0.0 &-999\\
6        &-1.0             & 0 &0& 22&2.814   & 2.648& 2.911  &2.371& 2.964& 88.162 &119 &-99.0 &0.0 &-999\\
11      &-1.0                   & 0 &0 &23&1.491 &1.352   & 1.672  &1.287 &1.768 &92.664 &115 &-99.0& 0.0 &-999\\
16      &0.280858             &3 &4 &24&0.328  &0.312  & 0.361 &0.302 &0.388 &99.442 &107 &-99.0 &0.0 &-999\\
17      & 0.768523         &2 &2& 22&0.689 &0.657   &0.723 &0.635 &0.811 &95.251 &120& -99.0& 0.0& -999\\

  \hline
\end{tabular}
\tablecomments{Excerpt from the photo-z catalog. Column 1: X-ray ID; Column 2 : spectroscopic redshift, when available; otherwise -1; Column 3: spectroscopic redshift quality flag. We consider reliable only redshifts for which this value is 3 or higher.  Column 4: spectroscopic redshift reference (0 = no z$_{spec}$; 1 = DEEP2+3; 2 = MMT (Coil et~al. 2009); 3 = CFRS (Lilly et~al, 1995); 4 = SDSS (DR9; Ahn et~al. 2012);  5 = LBG (Steidel et~al. 2003));  Column 5: Number of photometric points available for the fit ; Column 6: Photometric redshift; Column 7 and 8: 1$\sigma$ Lower and Upper value of photometric redshift;  Column 9 and 10: 3$\sigma$ Lower and Upper value of photometric redshift;  Column 11: [eak P(z); Column 12: Best fitting template: from 1 to 31 the templates are from S09, templates from 100+(1...30) are from the I09 library; Column 13, 14 and 15: as columns 6, 11,12 for the   second best photometric redshift when available, else -99, 0.0, -999.}
\end{center}
\end{table}
\end{rotate}


\end{document}